\documentclass[12pt]{article}
\usepackage[margin=0.95 in]{geometry} 
\usepackage{floatrow}
\usepackage{slashed}
\usepackage[all]{xy}
\usepackage{graphicx}
\usepackage{braket}
\usepackage{float}
\usepackage{mathtools}
\usepackage{mathrsfs} 
\usepackage[hidelinks]{hyperref}
\usepackage{cite}
\usepackage{tikz}
\usepackage{array, multirow}  
\numberwithin{equation}{section}

\def\bea{\begin{eqnarray}}
\def\eea{\end{eqnarray}}
\usepackage{amsmath,amssymb,amsthm,ytableau}
\usepackage{booktabs}
\usepackage{microtype}

\newcommand{\kk}{\Bbbk}

\newcommand{\End}{\mathrm{End}}
\newcommand{\Hur}{\mathsf{H}}
\newcommand{\Kar}{\mathrm{Kar}}
\newcommand{\R}{\mathbb{R}}

\usetikzlibrary{arrows.meta,decorations.markings,decorations.pathreplacing,decorations.pathmorphing}

\tikzset{
  ostrand/.style     = {semithick, -{Stealth[length=2.2mm,width=1.6mm]}},
  midarrow/.style    = {semithick,
                        decoration={markings,
                          mark=at position #1 with {\arrow{Stealth[length=2.2mm,width=1.6mm]}}},
                        postaction={decorate}},
  midarrow/.default  = 0.5,
  eqsign/.style      = {font=\large},
  khpic/.style       = {baseline=(current bounding box.center), scale=0.8},
  strand/.style={line width=0.9pt},
  uparrow/.style={postaction={decorate},decoration={markings,
     mark=at position #1 with {\arrow{Straight Barb[length=2.2pt,width=6pt,line width=0.9pt]}}}},
  uparrow/.default=0.5,
  perm box/.style={draw,line width=0.9pt,minimum width=1.7cm,minimum height=0.65cm,inner sep=0pt},
  kdot/.style={circle,fill,inner sep=0pt,minimum size=4.5pt},
  bulk/.pic={\draw[line width=0.9pt] (-2.8pt,-2.8pt) -- (2.8pt,2.8pt)
                                     (-2.8pt,2.8pt) -- (2.8pt,-2.8pt);}
}

\newcommand{\dcAup}[1]{(#1+0,0) .. controls (#1+0.85,0.55) and (#1+0.85,1.45) .. (#1+0,2)}
\newcommand{\dcAdown}[1]{(#1+0,2) .. controls (#1+0.85,1.45) and (#1+0.85,0.55) .. (#1+0,0)}
\newcommand{\dcBup}[1]{(#1+1,0) .. controls (#1+0.15,0.55) and (#1+0.15,1.45) .. (#1+1,2)}
\newcommand{\dcBdown}[1]{(#1+1,2) .. controls (#1+0.15,1.45) and (#1+0.15,0.55) .. (#1+1,0)}

\begin{document}

\hypersetup{pageanchor=false}
\begin{titlepage}
\vbox{\halign{#\hfil    \cr}}  
\vspace*{15mm}
\begin{center}
{\Large \bf Defects in Topological Symmetric Orbifolds}

\vspace*{2mm}

{\Large \bf and their String Dual}

\vspace*{10mm} 

{\large
Jan Troost}
\vspace*{8mm}

Laboratoire de Physique de l'\'Ecole Normale Sup\'erieure \\ 
 \hskip -.05cm
 CNRS, ENS, Universit\'e PSL, Sorbonne Universit\'e, \\
 Universit\'e Paris Cit\'e, 
 \hskip -.05cm F-75005 Paris, France	 

\vspace*{0.8cm}
\end{center}

\begin{abstract} {We review defect topological field theories with finite gauge groups. We then concentrate on restriction and induction defects between symmetric orbifold theories and recall that the defects form an algebra. This requires identifying the features of restriction and induction defects that are independent of the order of the symmetric group, which is natural in a grand canonical framework. The construction leads to Khovanov's categorification of the Heisenberg algebra. The bulk algebra of partial permutations corresponds to the endomorphisms of the identity defect in the category. We also interpret the  Hurwitz theory defects in the dual Gromov-Witten theory. This gives a defect descendant interpretation to higher level operators that arise in the integrable hierarchy of the Gromov-Witten theory of curves. In further applications, we relate known defects to our $S_n$ group theoretic description. For seamed surfaces, we determine the boundary operator algebra for higher values of the covering degree $n$ and provide asymptotics for the large $n$ growth of the number of boundary operators. 
  }  
\end{abstract}

\end{titlepage}

\hypersetup{pageanchor=true}

\tableofcontents

\section{Introduction} 
Two-dimensional topological quantum field theories with a gauge group equal to a finite group \cite{Dijkgraaf:1989pz,Fukuma:1993hy,Lazaroiu:2000rk,Moore:2006dw,Lauda:2006mn} underlie many symmetry structures in field theory. Defects between such theories permit a powerful abstract description \cite{Carqueville:2016nqk,Davydov:2011kb,Dougherty}. These defect topological field theories have an equally wide application range. In this paper, we mostly study their application to defects between topological symmetric orbifolds and their grand canonical counterparts. Concentrating on the symmetric groups $S_n$ and specific defects between the bulk theories provides us with extra structures. A grand canonical perspective on symmetric orbifolds has long been known to be useful in the closed theories (see e.g. \cite{Bantay:2000eq,Dijkgraaf:1996xw}), and this is strengthened in the $AdS_3/CFT_2$ correspondence where perturbative bulk string theory corresponds to the grand canonical ensemble on the boundary (see e.g. \cite{Aharony:2024fid}) in which we sum over all positive integer values of $n$. 

In the closed theory, a central problem is how to describe $n$-independent properties of the algebra of conjugacy class sums of $S_n$. Thus, the theory of the Farahat-Higman ring \cite{Farahat}, partial permutations \cite{IK} and symmetric function theory \cite{OO} became pertinent to describing physical structures like the bulk chiral ring structure constants \cite{Li:2020zwo}. The grand canonical perspective on the defect theory encourages us to also identify $n$-independent properties of defects between $S_n$ theories for various values of $n$.  
We will find that this point of view is again closely related to an interesting development in mathematics.
By studying the induction and restriction defects between $S_n$ and $S_{n \pm 1}$ theories, we will abstract properties that are independent of $n$ and identify the corresponding defect category. This construction is Khovanov's categorification of the Heisenberg algebra \cite{Khovanov}. Indeed, one can think of the demand of $n$-independence as corresponding to the requirement that one obtains an algebra after decategorification. Our insight into the defect algebra then leads us to conjecture how to integrate defects into the Gromov-Witten/Hurwitz correspondence \cite{OP}, which is a full-fledged gravity/gauge theory duality.

We start out in section \ref{DefectTQFT} by reviewing abstract properties of two-dimensional defect topological quantum field theories and how they apply to bulk two-dimensional theories with a finite group symmetry. 
We recall the Heisenberg categorification, its algebra and find its relation to the Hurwitz topological quantum field theory. We then discuss the topological defect theory and how it links up the mathematical and physics literature in section \ref{GrandCanonicalDefectTheory}. In particular, we argue that the $n$-independence necessary from the grand canonical perspective leads to a monoidal category. In section \ref{DefectsGWH} we give a proposal for how to extend the Gromov-Witten/Hurwitz correspondence to include defects and show how they naturally lead to defect descendant operators that play the leading role in the integrable structure of the theory. Thus, the operators introduced in the context of the equivariant calculation of bulk correlators \cite{Okounkov:2002cz} obtain a natural interpretation in the non-equivariant defect theory.  
In section \ref{OtherApplications}, we discuss other geometric applications of the defect topological symmetric orbifolds. Firstly, we make the link to the description of known defects in symmetric orbifold theories \cite{Harris:2025wak,HST} and secondly, to the construction of seamed surfaces \cite{Alexeevski:2002rp,AN1,Alexeevski:2007js}. We conclude in section \ref{Conclusions}.
\section{The Defect and Open-Closed Theories}
\label{DefectTQFT}
We discuss the defect topological quantum field theories for finite groups and their relation to category theory. We will work in the framework for defect theories explained in \cite{Carqueville:2016nqk} and we will use results of \cite{Davydov:2011kb}.

\subsection{The Defect Topological Quantum Field Theory}

 The pedagogical lecture notes \cite{Carqueville:2016nqk} describe the extension of open-closed topological quantum field theories to include defects as well as the reduction of the defect theory to their open-closed counterparts. We summarize this reference very briefly and refer to it for details. We study functors $Z$ from the category of defect bordisms $\text{Bord}^{\text{def}}(\mathcal{D})$ to the category of vector spaces. Our bordisms consist of two-dimensional surfaces that interpolate between marked circles and these surfaces represent scattering amplitudes between incoming and outgoing states. We augment this standard picture of closed topological quantum field theories in two dimensions by including surfaces that contain defect lines. The two-dimensional (oriented) bulk regions have a label in the set ${\cal D}_2$ while the (oriented) line defects are drawn from a set ${\cal D}_1$. The defect lines can end on boundary circles or they can be closed. 

 If we restrict to the case of no defect lines and a single label in the set ${\cal D}_2$, we have a closed topological quantum field theory and find an equivalence to a commutative Frobenius algebra \cite{Kock}. 
  Open-closed theories are obtained by allowing a non-trivial phase and an empty phase in the set ${\cal D}_2$ \cite{Carqueville:2016nqk}. The allowed boundary conditions are the labels in ${\cal D}_1$ (which are associated to defects between the empty and the non-trivial bulk phase). A circular boundary with the trivial bulk label on one side of a boundary point and a non-trivial label at the other side, separated by elements in the set of boundary conditions ${\cal D}_1$, becomes an interval with boundary conditions after erasing pieces of boundaries with trivial ${\cal D}_2$ label \cite{Carqueville:2016nqk}. The open-closed theories correspond to Calabi-Yau categories \cite{Lazaroiu:2000rk,Moore:2006dw,Lauda:2006mn}. The defect theory is richer than either of these two reductions. To explain its algebraic content, we briefly recall its correspondence to a 2-category \cite{Carqueville:2016nqk}. 

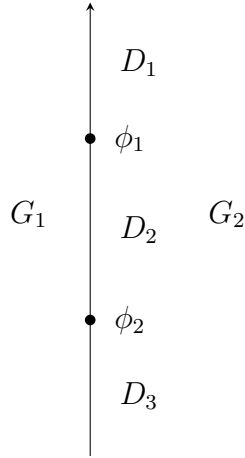
\begin{figure}[ht]
\begin{center}
\begin{tikzpicture}[>=stealth]

    \draw[->] (0,-3) -- (0,3);

    \fill (0,1.2) circle (2pt);
    \node[right=5pt] at (0,1.2) {$\phi_1$};

    \fill (0,-1.2) circle (2pt);
    \node[right=5pt] at (0,-1.2) {$\phi_2$};

    \node[right=7pt] at (0,2.2) {$D_1$};
    \node[right=7pt] at (0,0) {$D_2$};
    \node[right=7pt] at (0,-2.2) {$D_3$};

    \node[left=12pt] at (0,0.2) {$G_1$};
    \node[right=12pt] at (1,0.2) {$G_2$};

\end{tikzpicture}
\end{center}
\caption{The vertical composition of 2-morphisms}
\label{VerticalComposition}
   \end{figure}

\begin{figure}[ht]
   \begin{center}
   \begin{tikzpicture}

    \draw[->] (0,-1.2) -- (0,1.2);
    \draw[->] (3,-1.2) -- (3,1.2);

    \node at (0.5,0) {$D_1$};
    \node at (3.5,0) {$D_2$};

    \node at (-0.8,0.8) {$G_1$};
    \node at (1.5,0.8) {$G_2$};
    \node at (4.7,0.8) {$G_3$};

\end{tikzpicture}
\end{center}
\caption{The horizontal composition of 1-morphisms}
\label{HorizontalComposition}
\end{figure}
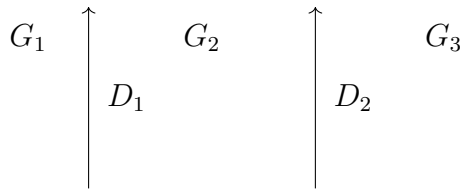

  The 2-category has bulk theories as objects, defects as 1-morphisms and defect junctions as 2-morphisms. We can compose 2-morphisms vertically
   \cite{Carqueville:2016nqk}. See Figure \ref{VerticalComposition} where we applied bulk labels $G_i$ to the left and the right of defects $D_i$ that meet at 2-morphism junctions $\phi_i$. 
   The horizontal composition of 1-morphisms or defects is graphically represented in Figure \ref{HorizontalComposition}. In physics, we interpret the vertical composition as an operator product of local operators, and horizontal composition as the fusion of defects. 

\begin{figure}[ht]
\begin{center}
 \begin{tikzpicture}[>=stealth, thick]

\begin{scope}[shift={(0,0)}]
    \draw[->]
        (-0.8,0)
        .. controls (-0.8,-1.2) and (0.8,-1.2)
        .. (0.8,0);
\end{scope}

\begin{scope}[shift={(3,0)}]
    \draw[<-]
        (-0.8,0)
        .. controls (-0.8,-1.2) and (0.8,-1.2)
        .. (0.8,0);
\end{scope}

\begin{scope}[shift={(6,0)}]
    \draw[->]
        (-0.8,0)
        .. controls (-0.8,1.2) and (0.8,1.2)
        .. (0.8,0);
\end{scope}

\begin{scope}[shift={(9,0)}]
    \draw[<-]
        (-0.8,0)
        .. controls (-0.8,1.2) and (0.8,1.2)
        .. (0.8,0);
\end{scope}

\end{tikzpicture}
\end{center}
\caption{Cups and caps}
\label{CupsAndCaps}
\end{figure}
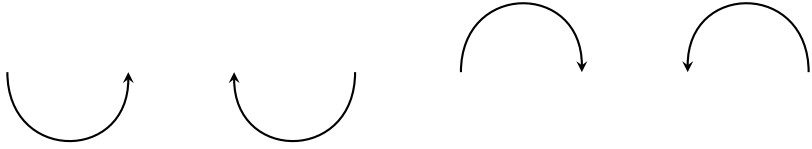

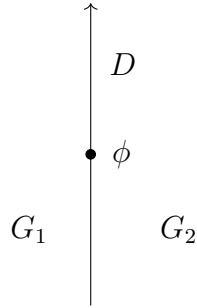
\begin{figure}[ht]
\begin{center}
\begin{tikzpicture}
    \draw[->] (0,-2) -- node[pos=0.8,right=3pt] {$D$} (0,2);

    \fill (0,0) circle (2pt) node[right=4pt] {$\phi$};

    \node[left=12pt] at (0,-1) {$G_1$};
    \node[right=22pt] at (0,-1) {$G_2$};
\end{tikzpicture}
\end{center}
\caption{A defect $D$ with 2-morphism $\phi$ between bulk theories $G_1$ and $G_2$.}
\label{GenericTwoMorphism}
\end{figure}

\begin{figure}[ht]
\begin{center}
\begin{tikzpicture}
    \draw[->,dotted] (0,-2) -- node[pos=0.8,right=3pt] {$1_G$} (0,2);
    \fill (0,0) circle (2pt) node[right=4pt] {$\phi$};

    \node[left=25pt] at (0,0.3) {$G$};
    \node at (0,1.5) {};
\end{tikzpicture}
\end{center}
\caption{For the identity defect, a 2-morphism gives a bulk operator. The identity defect is invisible, the bulk theory is $G$ everywhere and the 2-morphism is a point-like bulk operator $\phi$. }
\label{BulkOperatorTwoMorphism}
\end{figure}
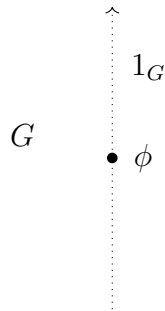

From the perspective of topological quantum field theories, in which we should be able to bend and straighten defects at will, it is necessary to restrict to 2-categories that have more structure. 
To obtain a defect topological theory, it is essential to also have a left and a right adjoint for each 1-morphism that are moreover equal. In diagrammatic language, this implies the existence of a cup and cap diagram of each orientation, as in Figure \ref{CupsAndCaps}. 
It is these cups and caps, and the consistency conditions they satisfy, that make the 2-category pivotal and that allow us to smoothly straighten out the topological defects (such that we can not only use progressive isotopies but all isotopies to establish the equivalence of topological defects).
The lecture notes
\cite{Carqueville:2016nqk} show that the algebra $ \End(1_G)$ of endomorphisms of the identity defect of a bulk theory $G$ is the commutative Frobenius algebra of operators of the corresponding bulk theory. Indeed, these endomorphisms are point-like objects living on an invisible line -- bulk operators. These are 2-morphisms since the unit $1_G$ is an invisible defect between identical bulk labels $G$ and an endomorphism can be thought of as a morphism from this defect to itself. See Figures \ref{GenericTwoMorphism} and \ref{BulkOperatorTwoMorphism}.
Many other structures and details are important -- we refer to \cite{Carqueville:2016nqk} and will mention in the following only those constructions that are crucial to us.

\subsection{Defects as Morphisms for Finite Group Gauge Theories}
We have discussed defect topological theories and their relation to 2-categories in general. In this subsection, we further specify these structures such that we can identify less general but more powerful properties of specific theories. In particular, for finite group gauge theories and their restriction and induction defects, stronger properties hold. 

\subsubsection{Separable Symmetric Frobenius Algebras}
A main example considered in \cite{Carqueville:2016nqk} and treated in detail in \cite{Davydov:2011kb} is relevant to us. It considers the category of separable symmetric Frobenius algebras $A$, with 1-morphisms given by their bimodules. The identity morphism $1_A$ is the algebra $A$ viewed as a bimodule over itself. Horizontal composition is the tensor product of bimodules. The left and right adjoints are equal to the dual bimodule. The field theory has a state sum construction \cite{Davydov:2011kb} such that we are within a standard field theory framework. We will work within an even smaller subcategory of this set-up. 

\subsubsection{Dijkgraaf-Witten Theories with Defects}
We concentrate on Frobenius algebras $\mathbb{C}[G]$ where $G$ is a finite group. The closed two-dimensional topological quantum field theories are (untwisted) Dijkgraaf-Witten theories \cite{Dijkgraaf:1989pz,Fukuma:1993hy}. The bulk observables live in the center $Z(\mathbb{C}[G])$ of the algebra $\mathbb{C}[G]$.\footnote{The normalization of the linear functional in the Frobenius algebra is chosen such that partition functions agree with the count of principal bundles.} 
The open-closed theories are specified by boundary conditions which in turn correspond to representations of the gauge group $G$ \cite{Lazaroiu:2000rk,Moore:2006dw,Lauda:2006mn}. 
The defect theory has defects between two bulk theories corresponding to groups $G_1$ and $G_2$. They are $(G_2,G_1)$ bimodules (and can be decomposed as a sum of products of irreducible representations of the two finite groups).
Two-morphisms are bimodule homomorphisms. 
This set-up gives a subcategory of the previous one. 

\subsubsection{The Restriction and Induction Defects}
We continue to shrink the set of objects under study. We concentrate on the defects that correspond to the restriction and induction functors of the finite representations of finite groups \cite{Khovanov} and their finite direct sums and tensor products. Thus, we study defects between bulk theories labeled by groups $H$ and $G$ for which $H \subset G$. A representation of the subgroup $H$ induces a representation of the group $G$ and a representation of $G$ can be restricted to be a representation of its subgroup $H$. These functors between representation categories uniquely correspond to the bimodules
 $(G)_H$ and ${}_H(G)$ where the notation indicates the module $\mathbb{C}[G]$ viewed either as a $G$ module on the left and an $H$ module on the right or vice versa. Thus, we study the induction-restriction subcategory of the category of group algebra bimodules. The 1-morphisms are compositions of induction and restriction functors in zigzags of inclusions. Properties of this subcategory are reviewed in \cite{Khovanov}. For groups, the induction and restriction functors are both left and right adjoint. This implies that there are
 four biadjointness 2-morphisms.
 These are precisely the cups and caps we drew before - intuitively, they relate a defect and a defect with its orientation reversed. They provide ways to annihilate or create pairs of defects. Despite the complexity of the formulation, these properties follow from elementary properties of groups and their subgroups as explained in \cite{Khovanov} in detail.
 Moreover, one has the property that endomorphisms on restriction and induction defect lines can be dragged along the line - a topological property necessary for our field theory. 
 This gives rise to (topological, deformable) intuitive string diagrams for the 2-category.  

For the purposes of \cite{Khovanov} and of this paper, an important extra property of the induction and restriction defects is the Mackey theorem that states, for subgroups $H,K \subset G$:
\begin{equation}
\mathrm{Res}^G_K \circ \mathrm{Ind}^G_H \cong \oplus_i \mathrm{Ind}^K_{K \cap g_i H g_i^{-1}} \circ \mathrm{Res}^{g_i H g_i^{-1}}_{K \cap \, g_i H g_i^{-1}}
\label{Mackey}
\end{equation}
where the sum is over coset representatives $g_i$ of $(K,H)$ double cosets of $G$, i.e. $G=\sqcup_i \, K g_i H$. This provides a way to change the order of restriction and induction functors. 
It gives an extra relation between the string diagrams  which is valid because we specified the structure to finite groups and induction and restriction defects. 

Let us summarize succinctly what we have reviewed so far \cite{Carqueville:2016nqk,Khovanov}. We have recalled that \cite{Khovanov} verified that the 2-category of restriction and induction defects for finite groups has the properties we expect of a defect topological quantum field theory \cite{Carqueville:2016nqk}. 
The theory has an extra property, which is the Mackey theorem (\ref{Mackey}). 

\subsection{The Symmetric Group Defect Algebra and Its Categorification}
\label{SymmetricGroupDefects}

We restrict even further the theories under scrutiny. We study only bulk groups that are symmetric groups $S_n$, and the induction and restriction functors associated to changing the value of $n$ by one \cite{Khovanov}.
This restriction allows for the introduction of extra interesting relations between diagrams.
In all diagrams, the region on the far right implicitly carries a bulk label $n$ for now and crossing an upward strand leftwards increases the label by one, while crossing the downward strand decreases it.
Crossings are allowed, and one finds relations between diagrams (via their corresponding bimodules) that match relations for the elements in the symmetric groups.
For instance, we have that \cite{Khovanov}:
\begin{center}
\begin{tikzpicture}[khpic]
  \draw[ostrand] \dcAup{0};
  \draw[ostrand] \dcBup{0};
  \node[eqsign] at (1.75,1) {$=$};
  \draw[ostrand] (2.5,0) -- (2.5,2);
  \draw[ostrand] (3.5,0) -- (3.5,2);
\end{tikzpicture}
\end{center}

\begin{center}
\begin{tikzpicture}[khpic]
  \draw[ostrand] (0,0) -- (2,2);
  \draw[ostrand] (2,0) -- (0,2);
  \draw[ostrand] (1,0) to[out=90,in=270] (1.6,1)
                       to[out=90,in=270] (1,2);
  \node[eqsign] at (2.75,1) {$=$};
  \draw[ostrand] (3.5,0) -- (5.5,2);
  \draw[ostrand] (5.5,0) -- (3.5,2);
  \draw[ostrand] (4.5,0) to[out=90,in=270] (3.9,1)
                         to[out=90,in=270] (4.5,2);
\end{tikzpicture}
\end{center}
These match the relations $s_{n+1}^2=1$ and $s_{n+2}s_{n+1}s_{n+2}=s_{n+1} s_{n+2} s_{n+1}$ in the symmetric group (where $s_n$ is the transposition $s_n=(n,n+1)$).
We moreover have the four equalities \cite{Khovanov}:

\begin{center}
\begin{tikzpicture}[khpic]
  \draw[ostrand] \dcAup{0};
  \draw[ostrand] \dcBdown{0};
  \node[eqsign] at (1.75,1) {$=$};
  \draw[ostrand] (2.5,0) -- (2.5,2);
  \draw[ostrand] (3.5,2) -- (3.5,0);
\end{tikzpicture}
\end{center}

\begin{center}
\begin{tikzpicture}[khpic]
  \draw[ostrand] \dcAdown{0};
  \draw[ostrand] \dcBup{0};
  \node[eqsign] at (1.75,1) {$=$};
  \draw[ostrand] (2.5,2) -- (2.5,0);
  \draw[ostrand] (3.5,0) -- (3.5,2);
  \node[eqsign] at (4.35,1) {$-$};
  \draw[ostrand] (5.1,2) .. controls (5.1,1.25) and (6.1,1.25) .. (6.1,2);
  \draw[ostrand] (6.1,0) .. controls (6.1,0.75) and (5.1,0.75) .. (5.1,0);
\end{tikzpicture}
\end{center}

\begin{center}
\begin{tikzpicture}[khpic]
  \draw[midarrow=0.52] (0,1) circle (0.75);
  \node[eqsign] at (1.6,1) {$= 1$};
\end{tikzpicture}
\end{center}

\begin{center}
\begin{tikzpicture}[khpic]
  \pgfmathsetmacro{\R}{0.4}
  \pgfmathsetmacro{\e}{0.25}
  \coordinate (X) at (0,1);
  \coordinate (O) at ({-sqrt(2)*\R},1);
  \draw[ostrand]
      (\e,0)
        .. controls (\e,0.5) and (\e,0.75) .. (X)
        -- ([shift={(45:\R)}]O)
        arc[start angle=45, end angle=315, radius=\R]
        -- (X)
        .. controls (\e,1.25) and (\e,1.5) .. (\e,2);
  \node[eqsign] at (1.05,1) {$= 0$};
\end{tikzpicture}
\end{center}
These four correspond to the Mackey decomposition (\ref{Mackey}) specified to the induction and restriction defects in the $S_n$ theories:
\begin{equation}
\mathrm{Res}_n^{n+1} \circ \mathrm{Ind}^{n+1}_n \cong \mathrm{Ind}^{n}_{n-1} \circ \mathrm{Res}_{n-1}^{n} \oplus \text{Id} \, .
\label{SymmetricMackey}
\end{equation}
Induction will always be with respect to fixing the last of the elements in $[n]=\{ 1,2,\dots,n \}$.
The Mackey relation (\ref{SymmetricMackey}) is also equivalent to the bimodule decomposition:\footnote{We use the notation $(n)_{n-1}(n)=\mathbb{C}[S_n] \otimes_{\mathbb{C}[S_{n-1}]} \mathbb{C}[S_n]$ et cetera. }
\begin{equation}
{}_n(n+1)_n \cong (n)_{n-1}(n)\oplus (n) \, .
\end{equation}
  The four relations follow from the fact that the direct sum decomposition with two terms has two associated injections and projections that imply these relations. An intuitive way to read the relation (\ref{SymmetricMackey}) is that we add a sheet (to the $n$ fold cover implicitly present in the $S_n$ fiber bundle over the surface) and then remove a sheet (on the left hand side of the relation). This is equivalent to removing a sheet  and then adding one, except for when we remove the sheet we add, which is represented by the identity term (on the right hand side). 
  Finally, we have a diagram which is an upward clockwise curl. It is non-trivial and we alternatively represent it by a dot on a upward oriented line. See Figure \ref{ClockwiseCurl}. When there are several clockwise curls, we can represent it with several dots or a number next to the dot \cite{Khovanov}.
  \begin{figure}[ht]
  \begin{center}
\[
\begin{tikzpicture}[baseline={([yshift=-.5ex]0,0)}]
  \draw[strand] (0,-1.1) -- (0,-0.6)
      to[out=90,in=225]  (0.12,0)     
      to[out=45,in=180]  (0.5,0.33)   
      to[out=0,in=90]    (0.82,0)     
      to[out=-90,in=0]   (0.5,-0.33)  
      to[out=180,in=-45] (0.12,0)     
      to[out=135,in=-90] (0,0.6);
  \draw[strand,uparrow=1] (0,0.6) -- (0,1.1);
\end{tikzpicture}
\quad = \quad
\begin{tikzpicture}[baseline={([yshift=-.5ex]0,0)}]
  \draw[strand,uparrow=1] (0,-1.1) -- (0,1.1);
  \node[kdot] at (0,0) {};
\end{tikzpicture}
\]
\end{center}
\caption{A clockwise curl decorates a defect with a dot.}
\label{ClockwiseCurl}
\end{figure}

Note that while we derived the relations by specifying a rightmost region to have label $n$, corresponding to a given symmetric group $S_n$, the diagrammatic relations are independent of $n$ \cite{Khovanov}. The unlabeled diagrams define Khovanov's category $H'$ \cite{Khovanov}. %
 Hence, the category $H'$ captures the properties of the induction and restriction functors in the category of $S_n$ representations in an $n$-independent manner. 

We can compare the category $H'$ to the category $S'$ which is the direct sum of the categories $S_k'$ of induction and restriction defects where the rightmost label is $k$. For each $k$ we  have a functor $F'_k : H' \rightarrow S_k'$ (by providing the label $k$ on the far right to each diagram in the category $H'$). Note that the category $S_k'$ does not have a monoidal structure. Arbitrary defects cannot be composed. They need to have the right neighbors in order for us to be able to draw them consecutively. By making abstraction of $n$, as we did, we obtain a monoidal structure on the category \cite{Khovanov}. By asking the question of how one describes defects grand canonically, one is rewarded with extra algebraic structure. These extra structures are also present in the 
Karoubi envelopes\footnote{In the Karoubi envelope of the category one adds idempotents. Here this means that we also consider the irreducible representations on which we can project the symmetric group action on consecutive restriction or induction functors via the idempotents $e_\lambda$ in the symmetric group algebra.} $H=\text{Kar}(H')$ and $S=\text{Kar}(S')$ of the categories $H'$ and $S'$. 

Each of those categories has an associated Grothendieck group. It is the monoidal categories $H'$ and $H$ that have a Grothendieck algebra. Physically, we may want to define an algebra of defect operators where we can multiply any of such operators. Clearly, this requires an $n$-independent description of the defects - a feat that is precisely accomplished in Khovanov's category $H'$. Finally, because of the Mackey relation (\ref{SymmetricMackey}), one finds
that the Grothendieck algebra of defects is the Heisenberg algebra \cite{Khovanov,BSW}. If we denote the induction functor $P=\mathrm{Ind}^{\cdot+1}_\cdot$ and the restriction functor $Q=\mathrm{Res}^{\cdot+1}_\cdot$ then their Grothendieck classes satisfy:
\begin{equation} [Q][P] - [P][Q] = 1 \, . \label{Heisenberg}
\end{equation}

\section{The Grand Canonical Defect Field Theory}

\label{GrandCanonicalDefectTheory}

In the previous section, we reviewed how appropriately defining the string diagrams for a defect theory between bulk patches with $S_n$ gauge groups and abstracting the structure universal in $n$ gave good motivation for Khovanov's categorification of the Heisenberg algebra \cite{Khovanov}.
To some extent, we now wish to follow the reverse road and we endeavor to understand how to derive properties of the field theory from the category. 
The lecture notes \cite{Carqueville:2016nqk} provide general guidance for this, but it is certainly worthwhile being explicit in this particular (infinite dimensional, intricate) example. We will see that there is a natural connection to the bulk Hurwitz topological quantum field theory describing the grand canonical theory of symmetric orbifolds \cite{OP,Benizri:2024mpx}, as well as to their boundary and defect counterparts.

\subsection{The Bulk Theory of Partial Permutations}

The bulk operators of the defect field theory corresponding to the category $H'$ should correspond to the  
 endomorphisms of the unit object $1$ in $H'$. These can be interpreted as bulk operators. This is in accord with the general construction of the closed theory in \cite{Carqueville:2016nqk}. 
For a monoidal category, the endomorphisms of the unit object form a commutative algebra. In the case of the category $H'$, there is an isomorphism $\varphi$ \cite{KLM}:
\begin{equation}
\varphi : \End_{H'}(1) \rightarrow \Lambda^\ast \, ,
\end{equation}
where $\Lambda^\ast$ is the algebra of shifted symmetric functions \cite{OO}. We conclude that the shifted symmetric function algebra $\Lambda^\ast$ is the algebra of bulk operators. This is precisely the algebra identified as the algebra of bulk operators in the Hurwitz or grand canonical symmetric orbifold theory \cite{OP,Benizri:2024mpx}. Thus, the bulk operators are sums over the orbits of partial permutations \cite{OP,Benizri:2024mpx,Li:2020zwo}, which become central conjugacy class sums after projection onto a given group algebra $\mathbb{C}[S_n]$.
It is satisfying that the $n$-independent formulation of the defect theories for $S_n$ leads to a category which rediscovers the Farahat-Higman/Ivanov-Kerov construction of the partial permutation algebra. The latter has been shown to be very useful not only in identifying the $n$-(in)dependence of the bulk structure constants \cite{Benizri:2024mpx,Li:2020zwo} but also in refining, proving and computing extremal correlators in $AdS_3/CFT_2$ \cite{Ashok:2023mow}. 

We also have the precise map $\varphi$ from closed 2-category string diagrams $\alpha_\mu$ to shifted symmetric polynomials evaluated at a partition $\lambda$ \cite{KLM}:
\begin{equation}
\varphi(\alpha_\mu)[\lambda] = \frac{(n \downharpoonright k)}{\text{dim} \, \lambda} \chi^\lambda(\mu \cup 1^{n-k})
\end{equation}
if $|\mu|=k \le n=|\lambda|$ and zero otherwise. 
The right hand side is a normalized character of the symmetric group \cite{IK,OO}. On the left, the closed diagram $\alpha_\mu$ in the category is the one where we have consecutive diagrams of cycles $\mu_i$ made up of $\mu_i$ upward moving strands with the last strand crossing over to the left and all other ones connecting vertically (shifted once towards the right). See Figures 
\ref{Strand} and \ref{Neighbors} -- or \cite{KLM}. Finally, the diagram is trivially closed into a circle towards the right \cite{KLM} to obtain an endomorphism of the unit object. 

\tikzset{
  strand/.style={line width=0.9pt},
  uparrow/.style={postaction={decorate},decoration={markings,
     mark=at position #1 with {\arrow{Straight Barb[length=2.2pt,width=6pt,line width=0.9pt]}}}},
  uparrow/.default=0.5,
  perm box/.style={draw,line width=0.9pt,minimum width=1.7cm,minimum height=0.65cm,inner sep=0pt},
}
\begin{figure}
\begin{center}
\begin{tikzpicture}[baseline={(0,0.33)}]
  \node[perm box] at (0.5,0.33) {\scriptsize $\mu_i$};
  \draw[strand,uparrow=0.56] (0,0.66) -- (0,1.45);
  \draw[strand,uparrow=0.56] (1,0.66) -- (1,1.45);
  \draw[strand] (0,0) -- (0,-0.75);
  \draw[strand] (1,0) -- (1,-0.75);
  \node at (0.5,1.22) {$\cdots$};
\end{tikzpicture}
\quad = \quad
\begin{tikzpicture}[baseline={(0,0.33)}]
  \draw[strand,uparrow=0.84] (0.35,-0.75) -- (0.35,1.45);
  \draw[strand,uparrow=0.84] (1.35,-0.75) -- (1.35,1.45);
  \node at (0.85,1.22) {$\cdots$};
  \draw[strand] (1.7,-0.75) -- (1.7,0.15)
      .. controls (1.7,0.5) and (1.55,0.55) .. (0.85,0.68)
      .. controls (0.05,0.83) and (0,0.85) .. (0,0.95);
  \draw[strand,uparrow=0.36] (0,0.9) -- (0,1.45);
  \draw[decorate,decoration={brace,amplitude=6pt}] (-0.1,1.6) -- (1.45,1.6)
      node[midway,above=6pt] {\footnotesize $\mu_i$ strands};
\end{tikzpicture}
\end{center}
\caption{Twisted strands corresponding to a $\mu_i$-cycle.}
\label{Strand}
\end{figure}
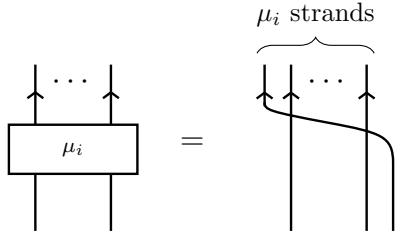
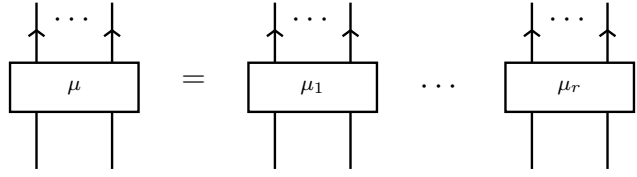
\begin{figure}
\begin{center}
\begin{tikzpicture}[baseline={(0,0.33)}]
\node[perm box] at (0.5,0.33) {\scriptsize $\mu$};
\draw[strand,uparrow=0.56] (0,0.66) -- (0,1.45);
\draw[strand,uparrow=0.56] (1,0.66) -- (1,1.45);
\draw[strand] (0,0) -- (0,-0.75);
\draw[strand] (1,0) -- (1,-0.75);
\node at (0.5,1.22) {$\cdots$};
\end{tikzpicture}
\quad = \quad
\begin{tikzpicture}[baseline={(0,0.33)}]
\node[perm box] at (0.5,0.33) {\scriptsize $\mu_1$};
\draw[strand,uparrow=0.56] (0,0.66) -- (0,1.45);
\draw[strand,uparrow=0.56] (1,0.66) -- (1,1.45);
\draw[strand] (0,0) -- (0,-0.75);
\draw[strand] (1,0) -- (1,-0.75);
\node at (0.5,1.22) {$\cdots$};
\node at (2.2,0.33) {$\cdots$};
\begin{scope}[xshift=3.4cm]
\node[perm box] at (0.5,0.33) {\scriptsize $\mu_r$};
\draw[strand,uparrow=0.56] (0,0.66) -- (0,1.45);
\draw[strand,uparrow=0.56] (1,0.66) -- (1,1.45);
\draw[strand] (0,0) -- (0,-0.75);
\draw[strand] (1,0) -- (1,-0.75);
\node at (0.5,1.22) {$\cdots$};
\end{scope}
\end{tikzpicture}
\end{center}
\caption{Neighboring twisted strands representing a conjugacy class $\mu$.}
\label{Neighbors}
\end{figure}

\subsection{The Open-Closed Theory}
Let us discuss how the open-closed theory for the grand canonical symmetric orbifold theory \cite{Troost:2025eqm} is contained in the defect theory as well. Note that the trivial representation of the trivial group $S_0$ induces the regular representation 
$P^n 1 \cong \mathbb{C}[S_n]$ after $n$ steps.\footnote{This can be proven step by step or by noting that induction is transitive.} Moreover, the regular representation contains every irreducible representation. By idempotent completion in the Karoubi envelope i.e. adding defects $(P^n,e_\lambda)$ where $e_\lambda$ is a (primitive) Young idempotent, one can project onto one irreducible representation (with multiplicity one). In this manner, we can reconstruct every possible boundary condition between an $S_n$ patch and an empty patch. One can realize this construction for each $n$ and therefore every possible irreducible representation of the partial permutation algebra can be realized with the defect theory we study. The other ingredients of the open-closed theory are then standard. 

\subsection{The Categorical View on the W-Algebra} There is a natural action of the annular diagrams in the category $H'$ on the closed diagrams (which we can think of as residing in a disk). It is obtained by drawing a new diagram which is a bigger disk consisting of the annular diagram surrounding the original disk diagram. The annular diagrams can be thought of as the trace of the category. It has been proven that this action is a $W_{1+\infty}$ action \cite{CLLS}. Thus, there is a $W_{1+\infty}$ action on $\Lambda^\ast$. To be more precise, we have a quotient of the $W_{1+\infty}$ algebra that acts. The ideal by which we divide corresponds to fixing the central charge of the W-algebra representation to one and the momentum to zero. In physical terms, the W-algebra is represented on a chiral boson of zero momentum. 

\section{The Defects in the Gromov-Witten/Hurwitz Duality}
\label{DefectsGWH}
In the previous sections, we have described how to usefully think about defects in the grand canonical symmetric orbifold. In this section, we address how to incorporate this insight into the calculation of correlators and indicate how one is guided towards an analogous construction in the dual Gromov-Witten theory (or topological string theory) on a curve. We do this by embedding our results of the previous sections into the oscillator calculations of \cite{OP} that connected localization calculations in Gromov-Witten theory \cite{Okounkov:2002cz} to Hurwitz numbers (see e.g. \cite{CavalieriMiles}).

We start out by providing a basic intuition. The induction operation $P$ corresponds to adding a sheet to the cover, while the restriction operation $Q$ corresponds to removing a single sheet from the cover, at the defect. 
There is therefore an analogy between the defect $P$ and the oscillator $\alpha_{-1}$ in the chiral boson Fock space formalism of \cite{OP} and the defect $Q$ and the oscillator $\alpha_{1}$. This analogy is immediately partially confirmed by the fact that these operators satisfy the Heisenberg relation. The repeated operation of induction $P^k$ can be modified by allowing for a twisting of the strands that we add. The choice of how we twist up the defect will lead to different (linear combinations of) $\alpha_{-\lambda}$ actions on the Fock space (with $|\lambda|=k$). Thus, apart from the jump in the covering degree, we must also specify a permutation profile of the jump. These intuitions are valid both on the Hurwitz side of the correspondence, and on the dual Gromov-Witten side. To render them quantitative, we very briefly review the bulk correspondence, which will allow us to introduce the necessary formalism to provide concrete formulas - for more background, see \cite{OP} and \cite{Benizri:2024mpx,Troost:2025eqm}. 

\subsection{A Brief Summary of the Bulk Correspondence}

For a closed target curve $X$ of genus $g$, the disconnected Hurwitz count $\Hur^\bullet_n$ of degree $n$ covers with insertions characterized by a ramification profile $\eta_i$ equals
\begin{equation}
\label{closed}
\Hur^\bullet_n (g;\{\eta_i\} )
=\sum_{\lambda\vdash n} (\frac{\dim\lambda}{n!} )^{2-2g}\prod_i\mathbf f_{\eta_i}(\lambda) \, ,
  \end{equation}
  where the $\mathbf f_\eta$ are properly normalized and generalized $S_n$ characters \cite{OP}. On the other hand, the disconnected stationary Gromov-Witten theory correlators are
 \begin{equation}
 \label{GWCorrelators}
  \langle\prod_i\tau_{k_i}(\omega)  \rangle^{\bullet X}_{n}
=\sum_{\lambda\vdash n} (\frac{\dim\lambda}{n!} )^{2-2g}\prod_i\mathcal{F}_{k_i+1}(\lambda) \, ,
\end{equation}
where the proper determination of the insertions $\mathcal{F}_{k}
=\overline{(k)}$ is central to the correspondence.\footnote{The normalization of the completed cycle $\overline{(k)}$ here is $1/(k-1)!$ times the one in \cite{OP}.} 
The precise statement \cite{OP} is that the bulk descendant operators $\tau_k(\omega)$ of the volume form $\omega$ on $X$ correspond to the completed cycles $\tau_k(\omega) \leftrightarrow \overline{(k{+}1)}$ where the latter have a very concrete algorithmic expansion in terms of orbits of partial permutations. After expanding the completed cycles, the correlators (\ref{GWCorrelators}) reduce to a Hurwitz count of the form (\ref{closed}) thus establishing the Gromov-Witten/Hurwitz correspondence. We must refer to the original paper \cite{OP} as well as the physical interpretation in \cite{Benizri:2024mpx} for much more background on this string theory / gauge theory correspondence. In this section, we concentrate on extending the correspondence to include defects.

\subsection{Defect Covers}

We start out with the geometric description of the generalized covers we consider when defects are present -- see \cite{HST} for a more extensive discussion. We then move to the realization of the defects as operators in a Fock space.  
\subsubsection{The Geometry of the Defect Covers}
We consider a closed (oriented) surface $X$ and a closed (oriented) one-dimensional defect curve $\Gamma \subset X$. The complement $X\setminus\Gamma $ is a union of disjoint regions $X\setminus\Gamma=\bigsqcup_c X_c$. Inside these regions, we can allow for a finite number of branch points in the set $B\subset X\setminus\Gamma$ which lie away from the defect curve. We will consider defect covers $\Sigma$ (possibly with boundaries) associated to maps $\pi :\Sigma \rightarrow X$ that are such that their boundaries map to the defect lines $\Gamma$, and that are branched covers of degrees $\{ n_c \}$ away from the defects. The map $\pi$ restricted to the boundary is a covering which is characterized by a winding for each boundary circle.

\begin{figure}
\begin{center}
\begin{tikzpicture}[scale=0.85,
  >={Stealth[length=2mm]},
  sheetT/.style   = {draw=black!70, fill=blue!6},
  sheetBm/.style  = {draw=black!70, fill=green!10},
  sheetBp/.style  = {draw=black!70, fill=orange!12},
  base/.style     = {draw=black, fill=black!4},
  defect/.style   = {very thick, red!80!black},
  defectback/.style = {thick, red!80!black, dashed},
  preim/.style    = {thick, red!70!black, densely dashed},
  bdryM/.style    = {line width=1.4pt, green!50!black},
  bdryP/.style    = {line width=1.4pt, orange!80!black},
  fibre/.style    = {black!45, densely dotted, thick},
  brace/.style    = {decorate, decoration={brace, amplitude=5pt}, thick},
  cut/.style      = {decorate, decoration={zigzag, amplitude=1.2pt, segment length=3pt}, thick, black},
]
\def\A{2}
\def\B{0.5}
\def\C{0.16}
\def\xf{-1.2}
\pgfmathsetmacro{\yf}{-\C*sqrt(1-(\xf/\A)^2)}
\def\xb{0.6}
\def\yBase{0} \def\yBm{1.9} \def\yTc{3.2} \def\yTb{4.5} \def\yTa{5.8}
\def\yBpb{7.1} \def\yBpa{8.4}

\newcommand{\Tsheet}[1]{%
  \draw[sheetT] (0,#1) ellipse ({\A} and {\B});
  \draw[preim] (\A,#1) arc (0:-180:{\A} and {\C});
  \draw[preim, black!35] (\A,#1) arc (0:180:{\A} and {\C});}
\newcommand{\Bmsheet}[1]{%
  \draw[sheetBm] (-\A,#1) arc (180:360:{\A} and {\B}) arc (0:180:{\A} and {\C});
  \draw[bdryM] (0,#1) ellipse ({\A} and {\C});}
\newcommand{\Bpsheet}[1]{%
  \draw[sheetBp] (-\A,#1) arc (180:0:{\A} and {\B}) arc (0:-180:{\A} and {\C});
  \draw[bdryP] (\A,#1) arc (0:-180:{\A} and {\C});
  \draw[bdryP, dashed] (\A,#1) arc (0:180:{\A} and {\C});
  \draw[cut] (\xb,#1+0.3) -- (\xb,{#1-\C*sqrt(1-(\xb/\A)^2)});
  \node[font=\small] at (\xb,#1+0.3) {$\times$};}

\draw[base] (0,\yBase) ellipse ({\A} and {\B});
\draw[defectback] (\A,\yBase) arc (0:180:{\A} and {\C});
\draw[defect, postaction={decorate, decoration={markings,
      mark=at position 0.62 with {\arrow{Stealth[length=2.2mm]}}}}]
     (\A,\yBase) arc (0:-180:{\A} and {\C});
\node[font=\small] at (\xb,\yBase+0.3) {$\times$};
\node[font=\scriptsize, right] at (\xb+0.02,\yBase+0.3) {$b$};
\node[font=\tiny] at (-0.3,\yBase+0.3) {$X_+$};
\node[font=\tiny] at (0.73,\yBase-0.34) {$X_-$};
\node[red!80!black, right] at (\A,\yBase-0.05) {$\gamma$};
\node[left] at (-\A-0.15,\yBase) {$X=S^2$};
\node[font=\scriptsize, align=left, right] at (\A+0.59,\yBase-0.05) {a defect line,\\ and $b\in B$ a branch point};

\Bmsheet{\yBm}
\Tsheet{\yTc} \Tsheet{\yTb} \Tsheet{\yTa}
\Bpsheet{\yBpb} \Bpsheet{\yBpa}

\foreach \y/\n in {\yBm/{b^-}, \yTc/{t_3}, \yTb/{t_2}, \yTa/{t_1}, \yBpb/{b^+_2}, \yBpa/{b^+_1}}
  \node[font=\small, left] at (-\A-0.05,\y) {$\n$};

\draw[<->, thick, orange!70!black] (\xb+0.12,\yBpa-0.17) to[bend left=35] (\xb+0.12,\yBpb+0.2);
\node[font=\scriptsize, orange!60!black, align=left, right] at (\xb+0.45,{(\yBpa+\yBpb)/2+0.1})
     {crossing the cut:\\ $b^+_1\leftrightarrow b^+_2$};

\draw[fibre] (\xf,\yBase+\yf) -- (\xf,\yBpa+0.75);
\foreach \y in {\yTc,\yTb,\yTa} \fill (\xf,\y+\yf) circle (1.6pt);
\fill[green!50!black]  (\xf,\yBm+\yf) circle (1.9pt);
\foreach \y in {\yBpb,\yBpa} \fill[orange!80!black] (\xf,\y+\yf) circle (1.9pt);
\fill[red!80!black] (\xf,\yBase+\yf) circle (1.9pt);
\node[font=\scriptsize, below] at (\xf,\yBase+\yf-0.02) {$x$};
\node[font=\scriptsize, fill=white, inner sep=1pt] at (\xf,\yBpa+0.95) {$\pi^{-1}(x)$};

\draw[->, thick] (-4.3,\yTb) -- (-4.3,\yBase+0.35) node[midway, left] {$\pi$};

\draw[brace] (-2.75,\yBm-0.3) -- (-2.75,\yBm+0.25)
   node[midway, left=6pt, font=\small, align=right] {$B_-$};
\draw[brace] (-2.75,\yTc-0.45) -- (-2.75,\yTa+0.45)
   node[midway, left=6pt, font=\small, align=right] {$T$};
\draw[brace] (-2.75,\yBpb-0.2) -- (-2.75,\yBpa+0.45)
   node[midway, left=6pt, font=\small, align=right] {$B_+$};

\draw[brace] (3.7,\yBpa+0.45) -- (3.7,\yTc-0.45)
   node[midway, right=6pt, font=\small, align=left]
   {$F_+=T\cup B_+$\\[1pt] $n_+
   =5$};
\draw[brace] (7,\yTa+0.45) -- (7,\yBm-0.3)
   node[midway, right=6pt, font=\small, align=left]
   {$F_-=T\cup B_-$\\[1pt] $n_-
   =4$};

\node[font=\scriptsize, align=left, right] at (2.45,\yBm)
   {a disk over $X_-$\\ with boundary winding $1$};
\node[font=\scriptsize, align=left, right] at (2.45,\yBpa+1.0)
   {a double cover of $X_+$\\
    branched at point $b$ and with boundary winding $2$};

\node[align=left, font=\small, anchor=north west] at (-3.9,-0.85)
 {$t=3,\qquad \nu^-_\gamma=(1),\qquad \nu^+_\gamma=(2) \, .
 $ 
   };
\end{tikzpicture}
\end{center}
\caption{A Defect Cover}
\label{DefectCover}
\end{figure}
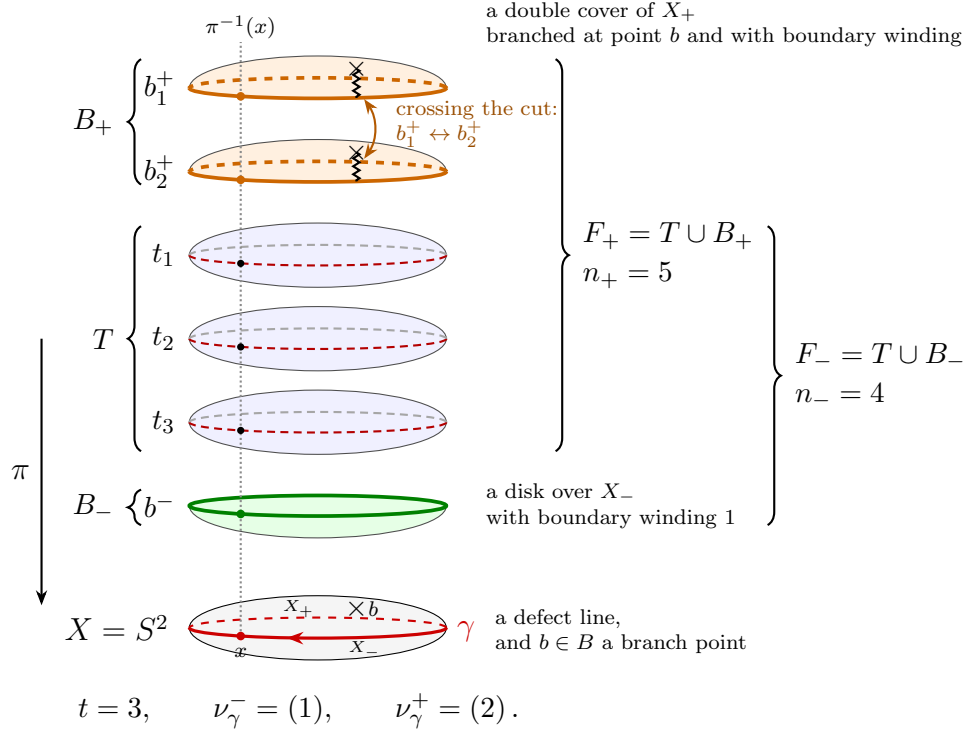

For concreteness, we consider one component $\gamma$ of the defect $\Gamma$ with regions $X_\mp$ on either side. Consider the inverse image $\pi^{-1}(x)$ of a point $x$ on the component $\gamma$. A preimage can be an interior point of $\Sigma$ with a neighborhood on both sides of $\gamma$, or a boundary point with a neighborhood that maps into $X_-$ or into $X_+$. We denote by $T,B_-,B_+$ the three respective subsets of preimages and we introduce the quantity $t=|T|$, namely, the number of continuing sheets. We also need the full set of preimages with a neighborhood on one or the other side, namely the sets:
\begin{equation}
F_-=T\cup B_- \, ,\qquad F_+=T\cup B_+ \, .
\end{equation}
The degrees of the left and right neighborhoods are therefore:
\begin{equation}
n_\mp=|F_\mp|=t+|B_\mp| \, .
\end{equation}
See Figure \ref{DefectCover} for an illustration.
If we turn once around the loop $\gamma$, we induce a permutation of the sheets on the left, and another permutation of the sheets on the right. We therefore find monodromies $g_\mp(\gamma)\in S_{n_\mp}$ acting on the set of preimages $F_\mp$. By the continuity of the sheets in $T$, the two permutations agree on this set. The sheets that are not transmitted must end somewhere -- they end on the defect. Their cycle types (or winding profiles) restricted to $B_\mp$, namely the type $\nu_\gamma^\mp$ of the monodromies $g_\mp(\gamma)$ enter into the degrees associated to the left and right bulk theories:
\begin{equation}
n_\mp=t+|\nu^\mp_\gamma| \,  \qquad \text{and} \qquad   n_+-n_-=|\nu^+_\gamma|-|\nu^-_\gamma| \, .
\end{equation}

\subsubsection{A Lightning Review of the Fock Space Formalism}
We wish to describe these defect covers concretely in the Fock space formalism of \cite{OP} which has been proven to be efficient in calculating large classes of correlators explicitly, as well as allowing for the exploitation of integrability in the bulk theory. 

In the paper \cite{OP}, two bases are used to describe covers and insertions. In the space of symmetric functions, these are the bases $p_\mu$ of power sums, and the Schur functions $s_\lambda$ directly related to characters of the symmetric group. 
The operators $\alpha_{-k}$ of the bosonic Fock space description of the covering surfaces are related to the power sum polynomials by:
\begin{equation}
\alpha_{-k}=p_k \, ,\qquad \alpha_k=k\,\partial_{p_k} \, ,
\qquad [\alpha_k,\alpha_{-l}]=k\,\delta_{k,l} \, .
\end{equation}
We can also regard the zero momentum bosonic Fock space as the zero charge infinite wedge fermionic states \cite{Ok,OP}, in which the Schur polynomial is related to the state with representation label $\lambda$, $|\lambda\rangle \cong s_\lambda$. We will make use of the operators \cite{Ok,OP}
\begin{equation}
  {\cal E}_r(z)=\sum_{k\in\mathbb{Z}+\frac12}e^{z(k-\frac r2)}E_{k-r,k} +\frac{\delta_{r,0}}{\zeta(z)} \, ,
\qquad \zeta(z)=e^{z/2}-e^{-z/2}=2\sinh\frac z2 \, ,
\end{equation}
where $E_{k-r,k}$ is an operator bilinear in the fermions. We recall the equations $\alpha_r=  {\cal E}_r(0)$ for $r\ne0$ and
\begin{equation}
\label{E0diag}
  {\cal E}_0(z) |\lambda\rangle = ( \frac{1}{\zeta(z)}+\sum_{i \ge 1} ( e^{z(\lambda_i-i+\frac12)}-e^{z(-i+\frac12)})) |\lambda\rangle = ( \frac{1}{z}+\sum_{k \ge 1}\frac{z^k}{k!}\mathbf{p}_k(\lambda)) |\lambda\rangle \, ,
\end{equation}
where $\mathbf{p}_k$ denotes the shifted symmetric functions of \cite{OO,OP} which are the symmetric function incarnations of the completed cycles \cite{OP}.

\subsubsection{Cutting in the Defect Operators}
The Fock space picture is rather intuitive. If we generate a sheet with a particular winding monodromy profile $\lambda$, it corresponds to the action of the oscillators $\alpha_{-\lambda}$, up to normalization. From this picture, we propose a natural guess for basic defect operators, which we then extend to more complicated cases.

Consider the induction defect $P$, inserted on a cycle that separates the cover into two manifolds. On the one side, we have $n_-$ sheets with a given profile, and on the other side $n_-+1$ sheets where the added sheet does not mix into the profile. In other words, it has (full) winding profile $\nu^+=\nu^- \cup \{ 1 \}$. Thus, it corresponds to inserting the action of $\alpha_{-1}$ into the Fock space representation of the correlator (which exists due to the assumption that the defect is separating). It is equally natural to claim that the word $P^a Q^b$ corresponds to the (extra) profiles $\nu^+=(1^a)$ and $\nu^{-}=(1^b)$ where we add or subtract the appropriate number of trivial sheets. Clearly, the relation $[ \alpha_{1}, \alpha_{-1}] = 1$ is consistent with the Heisenberg relation (\ref{Heisenberg}) inside the Grothendieck algebra associated to the Khovanov category.

A more general separating defect line insertion $\mathcal{D}_{\nu^+,\nu^-}$ is:
\begin{equation}
\label{windingdefect}
\mathcal{D}_{\nu^+,\nu^-} =\alpha_{-\nu^+}\,\alpha_{\nu^-} = (\prod_i\alpha_{-\nu^+_i}) (\prod_j\alpha_{\nu^-_j})
\, ,
\end{equation}
which maps an $n_-$ sheeted cover on one side to an $n_-+|\nu^+|-|\nu^-|$ sheeted cover on the other. The winding profile of the added sheets is captured by the partition of the creation or annihilation operators.\footnote{The precise normalization of the operators will depend on whether we label sheets. Here, we consider them labeled and omit exchange factors.}

\subsubsection{Twisted Strands and Young Projectors}
Let us be more precise about how to relate the Khovanov category picture to the winding profile. In the Karoubi envelope $\Kar({H}')$ one has the objects $P_\lambda=( P^{\,|\lambda|},e_\lambda)$, where $e_\lambda$ is a Young idempotent in the algebra $\mathbb{C}[ S_{|\lambda|}]$ which acts as endomorphisms on $P^{|\lambda|}$ (as demonstrated by the symmetric group relations satisfied by the defects, discussed in subsection \ref{SymmetricGroupDefects}). Namely, we have $|\lambda|$ upward strands on which the permutation group acts by braiding, and we can identify an idempotent in the algebra, associated to an irreducible representation $\lambda$. In the Grothendieck group, we have the translation of classes:
\begin{equation}
  [P_\lambda] \cong 
\sum_\mu z_\mu^{-1} \chi^\lambda(\mu) \, \alpha_{-\mu} \, ,
\end{equation}
where we move between the representation and the winding basis using the symmetric group characters. The objects of the Heisenberg category link to the representation basis of the Hilbert space, while the covers naturally provide the winding basis.  
A single long string, for instance, is related to an alternating sum of hooks (because of the Murnaghan-Nakayama rule for characters of the symmetric group):
\begin{equation}
\alpha_{-k} =p_k =\sum_{j=0}^{k-1}(-1)^j\, [P_{(k-j,\,1^j)}]
\, .
\end{equation}

\subsubsection{A Remark on the Gromov-Witten Framework}
From the Gromov-Witten side, the prescription we proposed above for the defect insertions must be derived from moduli space integrals. The defect corresponds to an open Gromov-Witten theory characterized by bulk and boundary cohomology classes. We expect degeneration theorems applied to separating defect lines to give the picture we intuitively argued above. 
In the bulk, the usual degeneration formula \cite{Okounkov:2002cz} glues with a partition of the unit operator thus enforcing equality of degrees. A defect line rather corresponds to a cut decorated by a Fock space operator. The sheets that are not matched across the cut correspond to the boundary circles of $\Sigma$ that end on the defect (on either side). Thus, we expect to be able to write the full amplitude as a product over bulk insertion contributions for each patch, multiplied by defect insertions between the various patches. The localization and degeneration techniques of \cite{Okounkov:2002cz} should be very useful to prove this picture. A natural first step is the equivariant localization calculation of Gromov-Witten invariants in the presence of an equatorial defect on $\mathbb{P}^1$, in analogy to, but beyond \cite{Okounkov:2002cz}.

\subsection{Dots and Descendants}
In this subsection, we move on to study  defects decorated by dots. Among other things, they will be useful in understanding how bulk operators can be moved through defects, which leads to the generation of a large W-algebra of defect (descendant) operators. 
\subsubsection{The Defect Descendants}
As argued above, the coefficient of $z^0$ of ${\cal E}_{-1}(z)$ is the defect $P$ that corresponds to the $\alpha_{-1}$ insertion. In this subsection, we analyze the role of the further terms in the operator:
\begin{equation}
  {\cal E}_{-1}(z)=\sum_{k\in\mathbb{Z}+\frac12}e^{z(k+\frac 12)}E_{k+1,k} \, .
\end{equation}
Firstly, we derive its action on the representation basis of states $|\lambda \rangle$ 
\begin{equation}
\label{Target}
  {\cal E}_{-1}(z)\, |\lambda\rangle= \sum_{\square\in\lambda^+} e^{z c(\square)} \, |\lambda+\square\rangle \, ,
\end{equation}
where $\lambda^+$ is the set of boxes that can be added to the Young diagram of $\lambda$, $\lambda+\square$ is the diagram with the box $\square$ added, and $c(\square)=j-i$ is the content of a box in row $i$ and column $j$. 

To derive this expression, we recall more details of the formalism of \cite{OP}. It provides a mathematical description of a fermion filled Dirac sea. 
For a subset $S=\{s_1>s_2>s_3>\cdots\}\subset\mathbb{Z}+\frac12$ such that both $S\setminus(\mathbb{Z}+\frac12)_{<0}$ and $(\mathbb{Z}+\frac12)_{<0}\setminus S$ are finite, we define the vector $v_S$:
\begin{equation}
v_S=\underline{s_1}\wedge\underline{s_2}\wedge\underline{s_3}\wedge\cdots \, .
\end{equation}
These vectors span the Hilbert space. The finiteness conditions ensure that we only excite a finite number of particles and holes above the filled Dirac sea. 
The vacuum is the state $v_\emptyset=\underline{-\frac12}\wedge\underline{-\frac32}\wedge\underline{-\frac52}\wedge\cdots$, with $S=(\mathbb{Z}+\frac12)_{<0}$ and charge zero. To relate this standard space to partitions, we associate to a partition $\lambda=(\lambda_1\ge\lambda_2\ge\cdots\ge0)$ the set 
\begin{equation}
S(\lambda)= \{\,s_i=\lambda_i-i+\frac12 \, |i=1,2,3,\dots\} \, ,
 \end{equation}
 and therefore the state
 \begin{equation}
v_\lambda=\underline{\lambda_1- \frac12} \wedge\underline {\lambda_2-\frac32} \wedge \underline{\lambda_3-\frac52} \wedge\cdots \, .
\end{equation}
We can identify this charge zero sector with the ring of symmetric functions, $v_\lambda\leftrightarrow s_\lambda$.
 
It is useful to recall a few standard pictures of the states in the infinite wedge space -- see Figure \ref{RussianDictionary} for an example. Firstly, we can picture $S(\lambda)$ as a Maya diagram, namely a row of positions $k\in\mathbb{Z}+\frac12$, each of which is occupied by a particle ($k\in S$) or a hole ($k\notin S$). Secondly, the relation to the Young diagram is through its profile in the Russian convention, in which the box in row $i$ and column $j$ is centered above the content point $x=j-i=c(\square)$. The boundary of the diagram is a path made of steps, one above each interval $[k-\frac12,k+\frac12]$. When $k$ is in the set $S(\lambda)$ we have that the step between $k-1/2$ and $k+1/2$ descends and vice versa for $k$ not in $ S(\lambda)$.
\tikzset{
  ry/wall/.style     = {black!25},
  ry/box/.style      = {fill=blue!8, draw=black!55},
  ry/newbox/.style   = {fill=orange!40, draw=orange!70!black, thick},
  ry/addable/.style  = {draw=orange!70!black, dashed},
  ry/profile/.style  = {very thick, line join=round},
  ry/axis/.style     = {-{Stealth[length=1.6mm]}, black!60},
  ry/guide/.style    = {black!20, densely dotted},
  ry/hop/.style      = {-{Stealth[length=2mm]}, thick, orange!70!black},
  ry/nohop/.style    = {-{Stealth[length=2mm]}, thick, black!35},
}
\newcommand{\RBox}[3][ry/box]{%
  \draw[#1] ({#3-#2},{#2+#3-2}) -- ++(1,1) -- ++(-1,1) -- ++(-1,-1) -- cycle;}
\newcommand{\Pt}[2]{\fill (#1,#2) circle (2.4pt);}
\newcommand{\Ho}[2]{\draw[fill=white] (#1,#2) circle (2.4pt);}
\begin{figure}[ht]
\centering
\begin{tikzpicture}[scale=0.8]
  \draw[ry/wall] (-4.5,4.5) -- (0,0) -- (4.5,4.5);
  \RBox{1}{1} \RBox{1}{2} \RBox{2}{1}
  \RBox[ry/addable]{3}{1}
  \RBox[ry/addable]{2}{2}
  \RBox[ry/addable]{1}{3}
  \draw[ry/profile] (-4.5,4.5) -- (-2,2) -- (-1,3) -- (0,2) -- (1,3) -- (2,2) -- (4.5,4.5);
  \Pt{-3.5}{3.5} \Pt{-2.5}{2.5} \Ho{-1.5}{2.5} \Pt{-0.5}{2.5}
  \Ho{0.5}{2.5}  \Pt{1.5}{2.5}  \Ho{2.5}{2.5}  \Ho{3.5}{3.5}
  \draw[ry/axis] (-4.6,-0.6) -- (4.8,-0.6) node[right] {\small content $c$};
  \foreach \c in {-4,...,4} {\draw[black!60] (\c,-0.52) -- (\c,-0.68);
                             \node[below, black!60] at (\c,-0.65) {\scriptsize $\c$};}
  \foreach \k/\h in {-3.5/3.5,-2.5/2.5,-1.5/2.5,-0.5/2.5,0.5/2.5,1.5/2.5,2.5/2.5,3.5/3.5}
     \draw[ry/guide] (\k,\h-0.12) -- (\k,-1.55);
  \draw[black!40] (-4.6,-1.7) -- (4.6,-1.7);
  \Pt{-3.5}{-1.7} \Pt{-2.5}{-1.7} \Ho{-1.5}{-1.7} \Pt{-0.5}{-1.7}
  \Ho{0.5}{-1.7}  \Pt{1.5}{-1.7}  \Ho{2.5}{-1.7}  \Ho{3.5}{-1.7}
  \node[left]  at (-4.6,-1.7) {\small $\cdots$};
  \node[right] at ( 4.6,-1.7) {\small $\cdots$};
  \node[below] at (-3.0,-1.85) {\scriptsize all filled};
  \node[below] at ( 3.0,-1.85) {\scriptsize all empty};
  \node at (0,5.0) {$\lambda=(2,1)$};
  \node[left, align=right] at (-5.5,-1.7) {\small Maya\\[-2pt]\small diagram};
\end{tikzpicture}
\caption{The partition $\lambda=(2,1)$ in Russian convention. Each step of the profile carries a dot, which is filled on a descending step and open on an ascending step. Projecting the dots onto the horizontal axis gives the Maya diagram. The boxes that can be added are indicated by dashed lines.}
\label{RussianDictionary}
\end{figure}
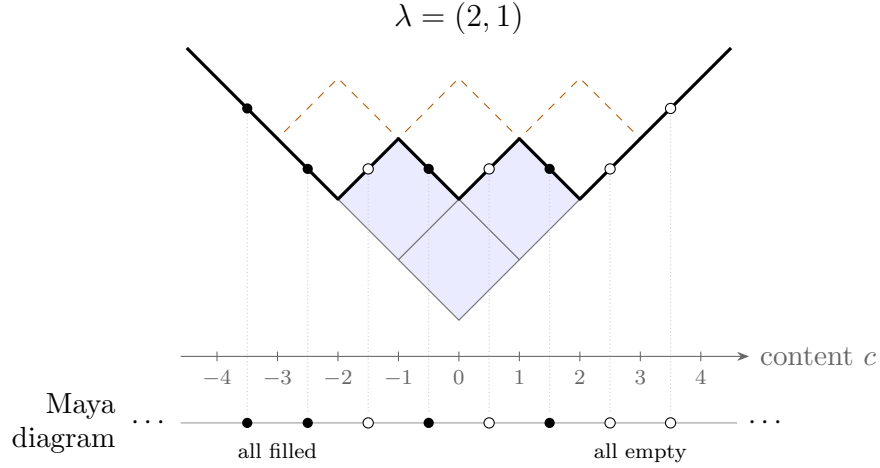

 A profile has a local minimum when the step on the left descends and the step on the right ascends. At such a local minimum, we can add a box and still have a valid Young diagram. 
Adding the box turns the descending and ascending pair of steps into an ascending and descending pair.  The corresponding particle in the Maya diagram moves one unit to the right. %

We define the fermion creation operator $\psi_k$ and the annihilation operator $\psi^*_k$ by
\begin{equation}
\psi_k\,v=\underline{k}\wedge v,
\qquad
\psi^*_k\, (\underline{s_1}\wedge\underline{s_2}\wedge\cdots)
=\sum_{j\ge1}(-1)^{j-1}\,\delta_{k,s_j} \,
\underline{s_1}\wedge\cdots\wedge\widehat{\underline{s_j}}\wedge\cdots \, ,
\end{equation}
where the hat denotes omission.
The operators satisfy the canonical anticommutation relation $\psi_i\psi^*_j+\psi^*_j\psi_i=\delta_{ij}$. The normal ordered fermion bilinears are
\begin{equation}
E_{ij}={:}\psi_i\psi^*_j{:}=
\begin{cases}
\psi_i\psi^*_j, & j>0,\\ -\psi^*_j\psi_i, & j<0 \, ,
\end{cases}
\end{equation}
while the operator ${\cal E}_{-1}(z)$ in terms of the elementary fermions reads:
\begin{equation}
\label{Em1}
  {\cal E}_{-1}(z)=\sum_{k\in\mathbb{Z}+\frac12}e^{z(k+\frac12)}\,\psi_{k+1}\,\psi^*_k\, .
\end{equation}
Each term therefore removes the particle at position $k$ and inserts a particle at position ${k+1}$, namely it moves a particle one unit to the right. This is only possible when the position $k$ is occupied and the position $k+1$ is empty. Otherwise, the result is zero.
Clearly, this coincides with the condition that a box of content $c=k+\frac12$ can be added to the Young diagram. 
The sum (\ref{Em1}) acting on the vector $v_\lambda$ therefore has finitely many non-zero terms, one for each addable box.

The prefactor is determined by observing that $k=s_i=\lambda_i-i+\frac12$. We replace it by exciting the value $s_i+1$ which is equivalent to replacing $\lambda_i$ by $\lambda_i+1$, so the new set is $S'=S(\lambda+\square)$ with a box at the position $(i,\lambda_i+1)$. The prefactor in the operator (\ref{Em1}) is
\begin{equation}
e^{z(k+\frac12)}=e^{z(\lambda_i-i+1)} = e^{z \, c(\square)} \, ,
\end{equation}
since the content of the added box is $c(\square)=(\lambda_i+1)-i$.
Summing over all the non-vanishing terms,
\begin{equation}
  {\cal E}_{-1}(z)\,|\lambda\rangle=\sum_{i:\ \text{$(i,\lambda_i+1)$ addable}} e^{z(\lambda_i+1-i)}\,|\lambda+(i,\lambda_i+1)\rangle =\sum_{\square\in\lambda^+}e^{z\,c(\square)}\,|\lambda+\square\rangle \, ,
\end{equation}
we find equation (\ref{Target}). We illustrate the procedure in our elementary example in Figures \ref{RussianLocalRule} and \ref{RussianAction}.
One similarly proves:
\begin{equation}
  {\cal E}_{1}(z)|\lambda\rangle=\sum_{\square\in\lambda^-}e^{z\,c(\square)}|\lambda-\square\rangle \, ,
\end{equation}
where the sum is over removable boxes in the Young diagram $\lambda$.

\begin{figure}[ht]
\centering
\begin{tikzpicture}[scale=0.9]
  \begin{scope}
    \draw[ry/profile] (-1.6,1.6) -- (0,0) -- (1.6,1.6);
    \RBox[ry/addable]{1}{1}
    \Pt{-0.5}{0.5} \Ho{0.5}{0.5}
    \draw[ry/axis] (-1.4,-0.4) -- (1.5,-0.4);
    \draw[black!60] (0,-0.32) -- (0,-0.48) node[below, font=\scriptsize] {$c$};
    \draw[ry/guide] (0,0) -- (0,-1.5);
    \draw[black!40] (-1.4,-1.5) -- (1.4,-1.5);
    \Pt{-0.5}{-1.5} \Ho{0.5}{-1.5}
    \draw[ry/hop] (-0.5,-1.38) to[bend left=50] (0.5,-1.38);
    \node[below, font=\scriptsize] at (-0.5,-1.6) {$k$};
    \node[below, font=\scriptsize] at ( 0.5,-1.6) {$k{+}1$};
    \node at (0,2.3) {\small before};
  \end{scope}
  \draw[-{Stealth[length=2mm]}, thick] (2.0,0.8) -- (3.0,0.8)
        node[midway, above, font=\small] {$e^{z c}$};
  \begin{scope}[xshift=5cm]
    \RBox[ry/newbox]{1}{1}
    \draw[ry/profile] (-1.6,1.6) -- (-1,1) -- (0,2) -- (1,1) -- (1.6,1.6);
    \Ho{-0.5}{1.5} \Pt{0.5}{1.5}
    \draw[ry/axis] (-1.4,-0.4) -- (1.5,-0.4);
    \draw[black!60] (0,-0.32) -- (0,-0.48) node[below, font=\scriptsize] {$c=k+\frac12$};
    \draw[black!40] (-1.4,-1.5) -- (1.4,-1.5);
    \Ho{-0.5}{-1.5} \Pt{0.5}{-1.5}
    \node[below, font=\scriptsize] at (-0.5,-1.6) {$k$};
    \node[below, font=\scriptsize] at ( 0.5,-1.6) {$k{+}1$};
    \node at (0,2.3) {\small after};
  \end{scope}
  \begin{scope}[xshift=10cm]
    \draw[ry/profile] (-1.6,2.2) -- (1.6,-1.0);
    \Pt{-0.5}{1.1} \Pt{0.5}{0.1}
    \draw[black!40] (-1.4,-1.5) -- (1.4,-1.5);
    \Pt{-0.5}{-1.5} \Pt{0.5}{-1.5}
    \draw[ry/nohop] (-0.5,-1.38) to[bend left=50] (0.5,-1.38);
    \draw[thick, red!70!black] (-0.13,-1.05) -- (0.13,-1.31)  (-0.13,-1.31) -- (0.13,-1.05);
    \node[below, font=\scriptsize] at (-0.5,-1.6) {$k$};
    \node[below, font=\scriptsize] at ( 0.5,-1.6) {$k{+}1$};
    \node at (0,2.3) {\small forbidden};
    \node[below, font=\scriptsize, align=center] at (0,-2.0) {target occupied:\\ no local minimum};
  \end{scope}
\end{tikzpicture}
\caption{The local rule. The term $e^{z(k+\frac12)}\psi_{k+1}\psi^*_k$ of ${\cal E}_{-1}(z)$ moves a particle from $k$ to $k+1$. If the position $k+1$ is empty, the profile has a local minimum at $c=k+\frac12$ and the move adds a box of content $c$ with weight $e^{zc}$. If the position $k+1$ is occupied, the term vanishes.}
\label{RussianLocalRule}
\end{figure}
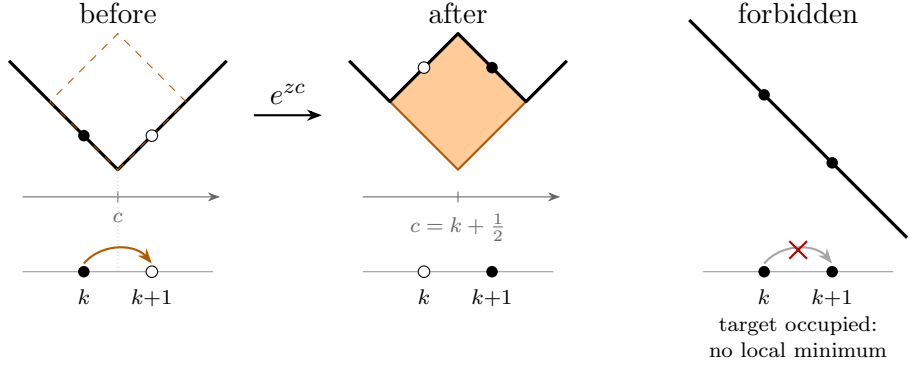

\begin{figure}[ht]
\centering
\begin{tikzpicture}[scale=0.8]
  \draw[ry/wall] (-4.5,4.5) -- (0,0) -- (4.5,4.5);
  \RBox{1}{1} \RBox{1}{2} \RBox{2}{1}
  \RBox[ry/addable]{3}{1} \RBox[ry/addable]{2}{2} \RBox[ry/addable]{1}{3}
  \draw[ry/profile] (-4.5,4.5) -- (-2,2) -- (-1,3) -- (0,2) -- (1,3) -- (2,2) -- (4.5,4.5);
  \draw[ry/axis] (-4.6,-0.6) -- (4.8,-0.6) node[right] {\small $c$};
  \foreach \c in {-4,...,4} {\draw[black!60] (\c,-0.52) -- (\c,-0.68);
                             \node[below, black!60] at (\c,-0.65) {\scriptsize $\c$};}
  \foreach \c in {-2,0,2} \draw[ry/guide] (\c,2) -- (\c,-2.6);
  \draw[black!40] (-4.6,-2.6) -- (4.6,-2.6);
  \Pt{-3.5}{-2.6} \Pt{-2.5}{-2.6} \Ho{-1.5}{-2.6} \Pt{-0.5}{-2.6}
  \Ho{0.5}{-2.6}  \Pt{1.5}{-2.6}  \Ho{2.5}{-2.6}  \Ho{3.5}{-2.6}
  \node[left]  at (-4.6,-2.6) {\small $\cdots$};
  \node[right] at ( 4.6,-2.6) {\small $\cdots$};
  \foreach \k/\kk/\w in {-2.5/-1.5/{e^{-2z}}, -0.5/0.5/{1}, 1.5/2.5/{e^{2z}}}
     \draw[ry/hop] (\k,-2.45) to[bend left=50]
           node[above, font=\small, text=orange!70!black] {$\w$} (\kk,-2.45);
  \draw[ry/nohop] (-3.5,-2.75) to[bend right=50] (-2.5,-2.75);
  \node[below, font=\scriptsize, black!50] at (-3.0,-3.05) {blocked};
  \node at (0,5.0) {$\lambda=(2,1)$};
\end{tikzpicture}
 
\vspace{1.5em}
 
\[
  {\cal E}_{-1}(z)\,
  \begin{tikzpicture}[scale=0.32, baseline={([yshift=-0.6ex]current bounding box.center)}]
    \RBox{1}{1} \RBox{1}{2} \RBox{2}{1}
  \end{tikzpicture}
  \,= 
  e^{-2z}\,
  \begin{tikzpicture}[scale=0.32, baseline={([yshift=-0.6ex]current bounding box.center)}]
    \RBox{1}{1} \RBox{1}{2} \RBox{2}{1} \RBox[ry/newbox]{3}{1}
  \end{tikzpicture}
  +\,
  \begin{tikzpicture}[scale=0.32, baseline={([yshift=-0.6ex]current bounding box.center)}]
    \RBox{1}{1} \RBox{1}{2} \RBox{2}{1} \RBox[ry/newbox]{2}{2}
  \end{tikzpicture}
  +\,
  e^{2z}\,
  \begin{tikzpicture}[scale=0.32, baseline={([yshift=-0.6ex]current bounding box.center)}]
    \RBox{1}{1} \RBox{1}{2} \RBox{2}{1} \RBox[ry/newbox]{1}{3}
  \end{tikzpicture}
\]
\caption{The action of ${\cal E}_{-1}(z)$ on $\lambda=(2,1)$. There are three particles that can hop one step to the right, corresponding to the three addable boxes. }
\label{RussianAction}
\end{figure}
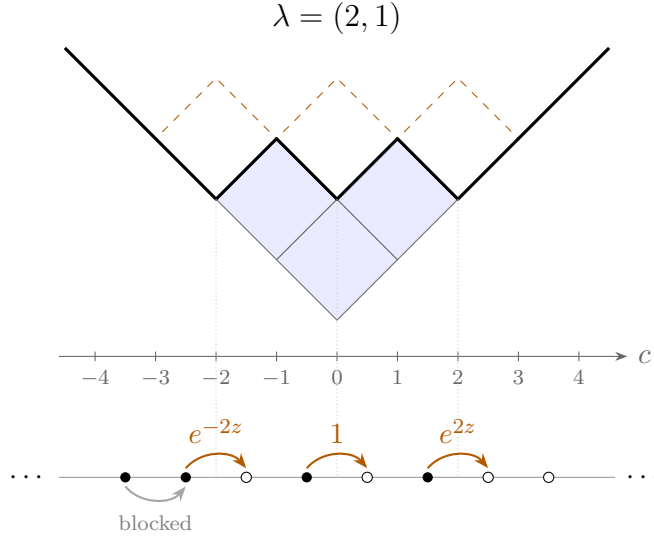

\subsubsection{Contents, Jucys-Murphy Elements and the Dotted Induction Defect}
 
In the previous subsection we derived that the operator $    {\cal E}_{-1}(z)$ acts by summing over each addable box $\square$ to the diagram $\lambda$ with weight $e^{z\,c(\square)}$. In this subsection we derive the same formula from induction and restriction for the symmetric groups, starting from decorated defects. We recall that in Khovanov's category $ H'$ the defect $P$ can be decorated by any number of clockwise curls, represented by dots on the $P$ strand. In summary, we will find that the dot acts as a Jucys-Murphy element \cite{Jucys,Murphy} on a state. Its eigenvalues are contents, and the generating function of dotted induction defects, with each $x$ representing one dot, namely $P\,e^{zx}=\sum_m\frac{z^m}{m!}P\,x^m$, acts on the Fock space as does the operator ${\cal E}_{-1}(z)$. Related calculations appear in \cite{LT,CGS}. Let us proof these claims. 

Firstly, we recall the Jucys-Murphy elements $L_k$ \cite{Jucys,Murphy} which are sums of transpositions:
\begin{equation}
L_k=\sum_{i=1}^{k-1}(i,k)
\, .
\end{equation}
It is not hard to show that the top element $L_{n+1}$ commutes with the (conjugation) action of the group $S_n$ that fixes element $n+1$. Therefore, the element $L_{n+1}$ lies in the centralizer of the subalgebra $\mathbb{C}[S_n]$ in $\mathbb{C}[S_{n+1}]$.  
Applying this for every $k$ shows that $L_1,\dots,L_{n+1}$ commute pairwise. Moreover the sum of all transpositions in $S_m$:
\begin{equation}
\label{sumJM} L_1+L_2+\cdots+L_{m}=\sum_{1\le i<j\le m}(i,j)=T_m \, ,
\end{equation}
is of course central in the group algebra $\mathbb{C}[S_m]$.
 
We recall that Young's branching rule states that the restriction of the representation $R^\Lambda$ associated to the Young diagram $\Lambda$ where $\Lambda\vdash n+1$, to $S_n$, is multiplicity-free:
\begin{equation}
\mathrm{Res}^{S_{n+1}}_{S_n}R^\Lambda\cong\bigoplus_{\square\in\Lambda^-}R^{\Lambda-\square} \, .
\end{equation}
Reciprocally, 
for $\lambda\vdash n$, we have that the induction
\begin{equation}
\mathrm{Ind}^{S_{n+1}}_{S_n}R^\lambda\cong\bigoplus_{\square\in\lambda^+}R^{\lambda+\square}
\end{equation}
is again multiplicity-free. 

 We want to establish that the operator $L_{n+1}$ acts on each branch of the induction by the content of the box. Let $\Lambda=\lambda+\square$ and consider the unique copy of $R^\lambda$ inside $\mathrm{Res} R^\Lambda$. Since the operator $L_{n+1}$ commutes with $S_n$, it maps each component of the restricted representation to itself, and by Schur's lemma and the fact that the representation is of multiplicity one, it acts on the copy of $R^\lambda$ by a scalar. To compute the scalar, we write $L_{n+1}=T_{n+1}-T_n$ using (\ref{sumJM}). Both terms act by scalars on this copy of $R^\lambda$. The full operator $T_{n+1}$ is central in $\mathbb{C}[S_{n+1}]$, so it acts on all of $R^\Lambda$ by its central character $f_{(2)}(\Lambda)$ and $T_n$ is central in $\mathbb{C}[S_n]$, so it acts on $R^\lambda$ by $f_{(2)}(\lambda)$. The central character of the class of transpositions is given by Frobenius' formula 
\begin{equation}
f_{(2)}(\mu) = \binom{|\mu|}{2} \frac{\chi^\mu ((1\,2))}{\dim\mu} = \sum_{\square\in\mu}c(\square) \, .
\end{equation}
The prefactor comes from counting transpositions. We conclude that
\begin{equation}
L_{n+1} |_{R^\lambda\subset\mathrm{Res} R^{\lambda+\square}} =f_{(2)}(\lambda+\square)-f_{(2)}(\lambda)=c(\square) \, .
\end{equation}
We can iterate this argument along inclusions $S_1 \subset S_2 \subset \dots \subset S_{n+1}$ and we find that at each step $L_k$ acts by the content of the added box.
Recall from section \ref{DefectTQFT} that the induction defect $P$ with right hand region labeled $n$ is the functor $\mathrm{Ind}^{S_{n+1}}_{S_n}=(n{+}1)_n\otimes_{\mathbb{C}[S_n]}-$. Moreover, we recall from \cite{Khovanov} that in the Khovanov category $ H'$ a dot on an upward strand is the clockwise curl and its image is the bimodule endomorphism
\begin{equation}
\rho_n:(n{+}1)_n\rightarrow(n{+}1)_n,\qquad g\mapsto 
g\,L_{n+1} \, .
\end{equation}
This is well defined on induced modules $\mathrm{Ind} \, V=\mathbb{C}[S_{n+1}]\otimes_{\mathbb{C}[S_n]}V$ because $L_{n+1}$ commutes with elements of $S_n$. 
We conclude that $\rho$ acts on the component $R^{\lambda+\square}$ by:
\begin{equation}
\rho |_{R^{\lambda+\square}\subset\mathrm{Ind} R^\lambda}=c(\square),
\qquad e^{z\rho} |_{R^{\lambda+\square}\subset\mathrm{Ind} R^\lambda}=e^{z\,c(\square)} \, .
\end{equation}
If we denote the defect decorated by dots in terms of a generating function, we have the equation:\footnote{To be slightly more precise: the defect class here indicates an element of the trace of the category, that acts on representations.}
\begin{equation}
[Pe^{zx}]\,|\lambda\rangle=\sum_{\square\in\lambda^+}e^{z\,c(\square)}\,|\lambda+\square\rangle =    {\cal E}_{-1}(z)\,|\lambda\rangle \, .
\end{equation}
Thus, the actions of ${\cal E}_{-1}(z)$ and the dot-decorated defect $P e^{zx}$ agree. This is what we wished to show. 
 Expanding in $z$, we find the identification:
\begin{equation}
[Px^m] = \sum_{k\in\mathbb{Z}+\frac12}
  (k+\frac12)^m
\, \psi_{k+1}\psi^*_k \, .
\end{equation}
We call the operators $P x^m$ defect descendants. Indeed, each dot inserts a clockwise curl on the newly added sheet, and we can think of the expansion in $z$ as a descendant expansion on the defect line. 
\subsubsection{Closure}
There is a simple argument for why we need to include the descendant operators into the operator algebra under study. When we pull an ordinary defect operator through bulk vertex operators, these leave behind descendant insertions on the defect. Let us show this.

\begin{figure}
\begin{center}
\[
\begin{tikzpicture}[baseline={([yshift=-.5ex]0,0)}]
  \pic at (-0.6,0) {bulk};
  \draw[strand,uparrow=0.864] (0,-1.1) -- (0,1.1);
\end{tikzpicture}
\quad = \quad
\begin{tikzpicture}[baseline={([yshift=-.5ex]0,0)}]
  \draw[strand,uparrow=0.864] (0,-1.1) -- (0,1.1);
  \pic at (0.6,0) {bulk};
\end{tikzpicture}
\quad + \quad
\begin{tikzpicture}[baseline={([yshift=-.5ex]0,0)}]
  \draw[strand,uparrow=0.864] (0,-1.1) -- (0,1.1);
  \node[kdot] at (0,0) {};
\end{tikzpicture}
\]
\end{center}
\caption{A sketch of the fact that a bulk vertex operator that switches sides of a defect creates new terms with dotted defects. The representation is symbolic since the coefficients and the number of dots on the right depend on the bulk vertex operator represented by a cross. The much more precise relation is equation (\ref{CommutationBulkAndDefect}).}
\label{Commutation}
\end{figure}
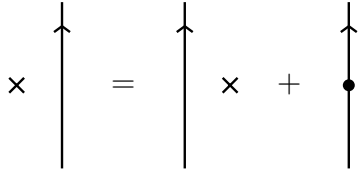

We start from the content identity for every partition $\lambda$,
\begin{equation}
\label{ContentId}
\zeta(z)\sum_{\square\in\lambda}e^{z\,c(\square)} = \sum_{i\ge1} (e^{z(\lambda_i-i+\frac12)}-e^{z(-i+\frac12)}) \, .
\end{equation}
To prove it, we note that both sides are sums over the rows of $\lambda$. The row $i$ contributes the contents $1-i,\dots,\lambda_i-i$ on the left hand side, and we have $\zeta(z)\sum_{j=1}^{\lambda_i}e^{z(j-i)}=e^{z(\lambda_i-i+\frac12)}-e^{z(-i+\frac12)}$, proving the formula.
Combining equation (\ref{ContentId}) with equation (\ref{E0diag}) gives the compact identity
\begin{equation}
  {\cal E}_0(z) =\frac{1}{\zeta(z)} + \zeta(z)\,\mathcal{C}(z) \, ,
\end{equation}
where we define the action of the operator $\mathcal{C}(z)$ on the representation basis by:
\begin{equation}
\mathcal{C}(z) | \lambda \rangle =\sum_{\square\in\lambda}e^{z\,c(\square)}  |\lambda \rangle \, .
\end{equation}
Now consider the same operator $\mathcal{C}_n(z)=\sum_{k=1}^n e^{zL_k}\in Z(\mathbb{C}[S_n])$ specified to live in a bulk labeled by $n$. The operator is central by Jucys' theorem
\cite{Jucys}, with eigenvalue $\sum_{\square\in\mu}e^{z\,c(\square)}$ on $R^\mu$ as previously argued. In the algebra $\mathbb{C}[S_{n+1}]$ one has
\begin{equation}
\label{telescope}
\mathcal{C}_{n+1}(z)=\mathcal{C}_n(z)+e^{zL_{n+1}} \, .
\end{equation}
Consider then a bulk operator $\mathcal{C}(z)$ placed in the region labeled $n+1$, to the left of a strand $P$. Therefore, on an induced module $\mathrm{Ind} \,  V$, we consider a bulk operator that acts by left multiplication by $\mathcal{C}_{n+1}(z)$, which, being central, equals right multiplication by it. A bulk operator in the region labeled $n$, to the right of the strand, acts on the module $V$ before induction, namely, by right multiplication by the operator $\mathcal{C}_n(z)$. By equation (\ref{telescope}) the difference is right multiplication by the extra term $e^{zL_{n+1}}$, that is,
\begin{equation}
\mathcal{C}(z)\, [P]- [P] \,\mathcal{C}(z)=[ P\,e^{zx} ]=    {\cal E}_{-1}(z) \, .
\label{CommutationBulkAndDefect}
\end{equation}
Identifying $    {\cal E}_0(z)=\zeta(z)^{-1}+\zeta(z)\,\mathcal{C}(z)$, we have reconstructed the commutator $[    {\cal E}_0(z),P]=\zeta(z)\,    {\cal E}_{-1}(z)$.
We conclude that moving a bulk operator through a parent defect $P$ gives rise to a dotted defect. See Figure \ref{Commutation} for a caricature of the commutation relation.

 We can repeat the whole story and establish a similar relation between the descendant restriction defect $Q e^{zx}$ and the operator ${\cal E}_{1}(z)$. The commutator 
\begin{equation}
  [\,Q\,e^{zx}, P\,e^{wx}\,] = [  {\cal E}_1(z),  {\cal E}_{-1}(w)] =\zeta(z+w)\,  {\cal E}_0(z+w) \, 
\end{equation}
can then be viewed as a version of the Mackey or Heisenberg relation which is dressed with all defect and bulk descendants. 

\subsubsection{Completed Cycles as Symmetric Functions of Jucys-Murphy Elements}
In passing, we want to point out the relation between completed cycles and Jucys-Murphy elements that we have implicitly established. We have the equation:
\begin{equation}
\mathcal{C}_n(z) 
=\sum_{m\ge0}\frac{z^m}{m!}\,p_m(L_1,\dots,L_n) \, ,
\end{equation}
where $p_m(L)$ is the $m$-th power sum of the Jucys-Murphy elements which is central \cite{Jucys,Murphy}. Expanding in $z$ and using $\zeta(z)=\sum_{j\ge0}z^{2j+1}/ (2^{2j}(2j+1)!)$ as well as comparing to equation (\ref{E0diag}), we can rewrite a completed cycle in terms of Jucys-Murphy elements:
\begin{equation}
\overline{(k)}=\sum_{0\le 2j\le k-1}\frac{p_{\,k-1-2j}(L)}{2^{2j}(2j+1)!\,(k-1-2j)!} + [z^k] (\frac{1}{\zeta(z)}-\frac1z) \, .
\end{equation}

\subsubsection{The Defect GW/Hurwitz Correspondence Revisited}
We are now in a position to slightly reformulate the defect GW/Hurwitz correspondence. We assume that every component of the defect curve $\Gamma$ in $X$ is separating, with winding profiles $\nu^\pm_\gamma$ on each defect component $\gamma$ and only stationary descendants in the bulk regions. Then we expect the disconnected open Gromov-Witten invariants of $(X,\Gamma)$ 
to be equal to the Hurwitz numbers of defect covers. Each bulk descendant class is to be replaced by a completed cycle. In the oscillator formalism, the defects are inserted as the operators (\ref{windingdefect}). Dotted defects correspond to the generating function $  {\cal E}_{\mp1}(z)$ which can be given a geometric interpretation through its realization as commutators of the elementary defect $P$ and the bulk operators $\tau_k(\omega)$. Again, the full geometric and moduli space analysis on the Gromov-Witten side requires more work.  

\subsection{Relations to Integrability and Further Remarks}

We end this section with a few comments.
\begin{itemize}
\item Closing the W-algebra: introducing the defects $P$ and $Q$ and closing their algebra with all bulk operators, we generated the defect descendant operators ${\cal E}_{\pm 1}$. If we close the algebra of these operators in turn, we generate all operators ${\cal E}_r$ that underlie the integrability and W-algebra of the solution of the Gromov-Witten/Hurwitz theory \cite{Okounkov:2002cz}. The W-algebra is expected to be the same as the one generated by the trace of the category $H'$ \cite{CLLS}. 
\item Since there are defects corresponding to the oscillators $\alpha_{\pm k}$, exponentiated defect insertions can be inserted in the grand canonical partition function as a coherent superposition.
This has as a consequence that inserting such an operator on a cut adjacent to a cap of $X=\mathbb{P}^1$ is the same as shifting the Toda times of the underlying $\tau$-function.
This is akin to open-closed string duality where we have a similar relation between brane insertions and shifted times, in particular in lower dimensional string theories.
\item We expect a  non-separating defect to give rise to a trace over the state space, with oscillator insertions.  
\item We expect to be able to extend the constructions above to a non-trivial topological seed theory by replacing the Khovanov category $H'$ with the Frobenius Heisenberg category of \cite{RS,Savage}.
\end{itemize}

\section{More Applications}
\label{OtherApplications}
In this section, we broaden our horizon and discuss a few further applications of the topological defect theories we reviewed in section \ref{DefectTQFT}. We concentrate on two applications that tie our approach to the literature. In the first, we study topological members of the set of defects discussed in \cite{HST}. We identify the relevant bimodules and how the construction fits into the Khovanov grand canonical approach. Secondly, we discuss the seamed surfaces of \cite{Alexeevski:2002rp,AN1,Alexeevski:2007js} in the general framework we provided. We algebraically compute extra data on the boundary operator algebras that were described diagrammatically in \cite{Alexeevski:2002rp,AN1,Alexeevski:2007js}, illustrating the power of the reformulation.

\subsection{Grand Canonical Transmission}
As a preliminary, we remark that from the perspective of symmetric orbifold conformal field theories, we have mostly concentrated on a topologically twisted seed theory with only an identity operator. The choice of a trivial seed theory implies that e.g. the boundary state in the seed theory is trivial and we can suppress all reference to the seed altogether. We moreover concentrate on topological defects, as before. 
\subsubsection{The D-branes}

We reviewed that a D-brane boundary state corresponds to a defect between the trivial theory and the symmetric orbifold theory in section \ref{DefectTQFT}. Comparing our set-up to the analysis of boundary states in symmetric orbifolds \cite{Belin:2021nck,Gaberdiel:2021kkp}, we note that we can define irreducible representation boundary states:
\begin{equation}
|{\lambda}\rangle_C
= \frac{1}{\sqrt{n!}} \sum_{ g \in S_n} {\chi^{\lambda} (g)} |g \rangle \, ,
\end{equation}
where $g$ represents a boundary holonomy. This is the standard translation of an orbifold boundary state from the holonomy basis to the representation basis.
It is a defect state (between the symmetric orbifold theory and the empty theory) labeled by a single representation $\lambda$ of $S_n$. Thus, the topological theory captures the symmetric group combinatorics of the physical boundary states.

\subsubsection{The Topological Defects}
We turn to (the topological counterparts of) the defects discussed in \cite{HST}. The paper \cite{HST} distinguishes $t$ trivially connected sheets in the cover above the interface, from the other $n_{\pm}-t$ sheets that dangle to the left and the right of the defect.  
Here, we wish to describe these defects from our bimodule perspective. To that end, we first note that the gauge groups on the left and on the right of the defect form the product $S_{n_+} \times S_{n_-}$. It is useful for us to identify the part of the gauge group that is shared between the left and the right theories due to the $t$ sheets that run through the defect. Thus, we identify a subgroup $H=S_{n_+-t} \times S_t \times S_{n_--t}$ where we diagonally embed the $S_t$ group in the two factors of $S_{n_+=|\nu^+|+t} \times S_{n_-=|\nu^-|+t}$. The idea is now clear: we fix a module of the subgroup $H$, i.e. three characters $\chi_{\pm},\chi_{t}$ that identify a representation $V$ of the subgroup $H$. We then propose that the relevant topological bimodule that captures the topology of the defects of \cite{HST} is equal to the induced bimodule $\mathcal D_{\chi_\pm,\chi_t}$:
\begin{align}
\mathcal D_{\chi_\pm,\chi_t} &= \mathrm{Ind}_{H}^{S_{n_+} \times S_{n_-}} (V_{\chi_{+}} \otimes V_{\chi_t}\otimes V_{\chi_-} )
\, .
\end{align}

\subsubsection{The Relation to the Elementary Defects}
To relate these defect bimodules to the  defects $P$ and $Q$, we can think of the inclusions
\begin{equation}
S_{n_\pm-t}\subset S_{n_\pm}
\end{equation}
as induction/restriction processes that can be resolved into braids of elementary $P$ and $Q$ interfaces. Schematically, the interface has the form
\begin{equation}
\mathcal D_{\chi_\pm,\chi_t}
\sim (P^{n_+ - t},e_{\chi_+}) \circ 
{\mathrm{Ind}}_{\Delta S_t,\chi_t}^{S_t\times S_t}(V_t)
\circ (Q^{n_--t},e_{\chi_-}) \, .
\end{equation}
The diagonal factor
\begin{equation}
{\mathrm{Ind}}_{\Delta S_t,\chi_t}^{S_t\times S_t}(V_t)
\end{equation}
is due to the fact that the $t$ transmitted colors are identified between the two sides.

The interface is labeled by three representations. The characters $\chi_-$ and $\chi_+$ play a standard role and project onto representations of the sheets that do not cross the defect, while the character $\chi_t$ lives on the transmitted sheets, which are shared by both sides.
To write it in terms of the elementary defects requires a bit more work and some extra insight. 
Suppose we have a representation $W$ of $S_t$ arriving at the $t$-sheeted transmissive defect. We can then apply the projection formula for restriction and induction to write:
\begin{equation}
\label{ProjectionFormula}
\mathrm{Ind}_{S_0}^{S_t}\,\mathrm{Res}^{S_t}_{S_0}\,W \cong W\otimes\mathbb{C}[S_t] \, .
\end{equation}
In Khovanov's category, the left hand side is the word $P^{t}Q^{t}$ placed in a region labeled $t$. The $t$ downward strands restrict all the way to the trivial group, such that the region between them and the upward strands is labeled $0$. The $t$ upward strands then induce back to $S_t$. Crossings of the upward strands give a copy of $S_t$ acting on $P^{t}$, and crossings of the downward strands give another copy acting on $Q^{t}$ \cite{Khovanov}. Consider a permutation $\sigma$ that acts by the crossing $\sigma$ on the upward strands and the crossing $\sigma^{-1}$ on the downward strands. This defines a diagonal action.
The diagonal action is trivial on $W$ and acts on the factor $\mathbb{C}[S_t]$ in equation (\ref{ProjectionFormula}) by right multiplication.

\tikzset{
  dg/strand/.style = {thick, -{Stealth[length=2mm]}},
  dg/pairA/.style  = {dg/strand, orange!80!black},
  dg/pairB/.style  = {dg/strand, blue!70!black},
  dg/pairC/.style  = {dg/strand, green!50!black},
  dg/mirror/.style = {black!40, dashed},
  dg/label/.style  = {font=\small},
  dg/box/.style    = {draw, fill=white, rounded corners=1pt, minimum height=6mm},
}
 
\begin{figure}[ht]
\centering
\begin{tikzpicture}[scale=0.85]
  \draw[dg/pairC] (0,0) -- (0,0.5) to[out=90,in=270] (1,1.5) to[out=90,in=270] (2,2.5) -- (2,3);
  \draw[dg/pairB] (1,0) -- (1,0.5) to[out=90,in=270] (0,1.5) -- (0,3);
  \draw[dg/pairA] (2,0) -- (2,1.5) to[out=90,in=270] (1,2.5) -- (1,3);
  \draw[dg/pairC] (4,3) -- (4,2.5) to[out=270,in=90] (5,1.5) to[out=270,in=90] (6,0.5) -- (6,0);
  \draw[dg/pairB] (6,3) -- (6,1.5) to[out=270,in=90] (5,0.5) -- (5,0);
  \draw[dg/pairA] (5,3) -- (5,2.5) to[out=270,in=90] (4,1.5) -- (4,0);
  \draw[dg/mirror] (3,-0.25) -- (3,3.25);
  \node[dg/label] at (-0.7,1.5) {$3$};
  \node[dg/label, fill=white, inner sep=1pt] at (3,1.5) {$0$};
  \node[dg/label] at (6.7,1.5) {$3$};
  \node[dg/label, below, align=center] at (1,-0.15) {$P^{\,3}$\\ crossings of $\sigma$};
  \node[dg/label, below, align=center] at (5,-0.15) {$Q^{\,3}$\\ mirror image};
  \node[dg/label] at (-0.7,3.4) {(a)};
\end{tikzpicture}
\hspace{2.5em}
\begin{tikzpicture}[scale=0.85]
  \foreach \x in {0,1,2} \draw[dg/strand] (\x,0) -- (\x,3);
  \foreach \x in {4,5,6} \draw[dg/strand] (\x,3) -- (\x,0);
  \node[dg/box, minimum width=2.4cm] (U) at (1,1.5) {$e_\lambda$};
  \node[dg/box, minimum width=2.4cm] (D) at (5,1.5) {$\overline{e_\lambda}$};
  \draw[dg/mirror] (U.east) -- node[above, dg/label, black] {$\Delta$} (D.west);
  \node[dg/label] at (-0.95,1.5) {$3$};
  \node[dg/label] at (3,0.4) {$0$};
  \node[dg/label] at (6.95,1.5) {$3$};
  \node[dg/label, below, align=center] at (3,-0.15) {$e^\Delta_\lambda$ on $P^{\,3}Q^{\,3}$\\ \phantom{x}};
  \node[dg/label] at (-0.95,3.4) {(b)};
\end{tikzpicture}
\caption{The diagonal action on $P^{t}Q^{t}$, drawn for $t=3$. (a) A permutation acts by its crossing diagram on the upward strands and by the mirror diagram on the downward strands. (b) Our abbreviation for the diagonal idempotent $e^\Delta_\lambda$, namely a Young idempotent on the upward strands together with its mirror image on the downward strands. This is a schematic representation because the dashed line in the middle hides a complicated decomposition.}
\end{figure}
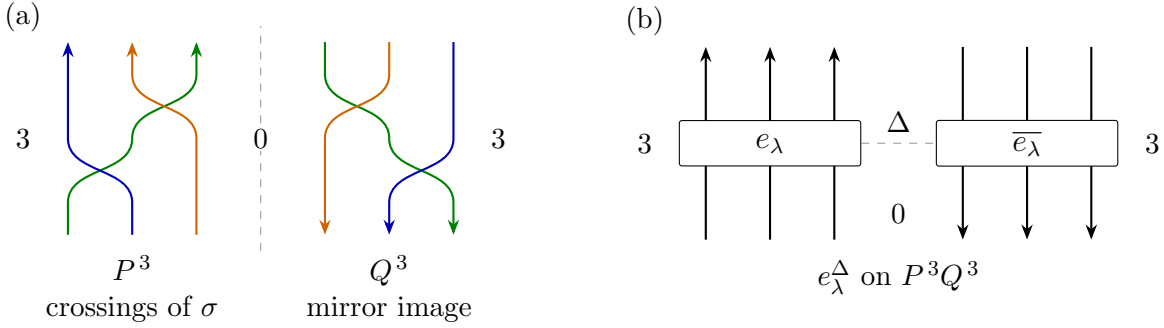

 The regular representation contains every irreducible representation, and the copy of $R^\lambda$ is cut out by right multiplication with a Young idempotent $e_\lambda$: $\mathbb{C}[S_t]\,e_\lambda\cong R^\lambda$. Applying $e_\lambda$ through the diagonal action, which we denote $e^\Delta_\lambda$, therefore gives (by the projection formula (\ref{ProjectionFormula})):
\begin{equation}
\label{KroneckerObject}
  (P^{t}Q^{t},\,e^\Delta_\lambda)   W \mapsto W\otimes R_\Delta^\lambda \, .
\end{equation}
Note that it requires more work to decompose the tensor product representation into a sum of tensor products of irreducibles.
 In summary, starting from a region labeled $n_-$, we restrict $n_--t$ times and project the corresponding downward strands with $e_{\chi_-}$. This removes the reflected sheets on the left and keeps only the part of the representation of type $\chi_-$ on them. We then restrict $t$ more times and induce $t$ times back, and we couple these strands with $e^\Delta_{\chi_t}$ in order to describe the transmitted sheets. Finally, we induce $n_+-t$ more times and project the new upward strands with $e_{\chi_+}$. This creates the reflected sheets on the right. 

Let us briefly remark on the grand canonical point of view. In Khovanov's category, the regions carry no labels, and a given diagram makes sense in a region regardless of its label -- the structure is independent of $n$. If the object (\ref{KroneckerObject}) is placed in a region labeled $n\ge t$ instead of $t$, we have $n-t$ extra sheets passing through (which give rise to extra combinatorial factors).
The $n-t$ extra sheets pass through without any projector, namely they are additional transmitted sheets, and the $n$-independent object is the interface with an arbitrary number of these extra sheets.

\subsubsection{Generalization}
 The topological theory captures the $S_n$ combinatorics of the defects. If we have other gluing data associated to the seed, then the appropriate category for grand canonical defects will again be a generalized Heisenberg category built on the boundary-and-defect  algebra of the seed \cite{RS,Savage} as mentioned at the end of section \ref{DefectsGWH}.

\subsection{On Seamed Surfaces}
\label{SeamedSurfaces}
In this subsection, we consider an open-closed theory with symmetric group symmetry that has drawn attention in the literature because of its geometric interpretation \cite{Alexeevski:2002rp,AN1,Alexeevski:2007js}. It was argued to generalize Hurwitz numbers in an interesting manner. The theory we consider is the theory of seamed surfaces. These are surfaces that one obtains by gluing ordinary surfaces at seams \cite{Alexeevski:2002rp,AN1,Alexeevski:2007js}. The counting of covers of surfaces by seamed surfaces maps onto the calculation of amplitudes in an open-closed topological field theory with symmetric gauge group in a manner that generalizes the field theory description of the Hurwitz count of covers of Riemann surfaces \cite{CavalieriMiles}. We want to understand the seamed surface theory in the canonical terms that we recalled in section \ref{DefectTQFT}. Since the bulk theory is a standard $S_n$ Dijkgraaf-Witten theory, this mostly entails determining the representation-theoretic content of the open boundary conditions that were identified to correspond to seamed surfaces \cite{Alexeevski:2002rp,AN1,Alexeevski:2007js}.

\subsubsection{Seamed Surfaces and a Symmetric Group Representation}
\begin{figure}
\begin{center}
\begin{tikzpicture}[v/.style={circle,fill,inner sep=1.8pt}]
  \foreach \i in {1,2,3} \node[v]  (a\i) at (0,{1.5-\i})   {};
  \foreach \j in {1,...,4} \node[v] (b\j) at (3,{2.5-\j}) {};
  \foreach \x/\y in {1/1, 1/3, 2/2, 1/4, 3/3} \draw[thick] (a\x) -- (b\y);
\end{tikzpicture}
\end{center}
\caption{The left vertices represent the boundary of a seamed surface which corresponds to a set partition $\{ \{1,2,3\} ,\{4 \}, \{5 \} \}$ of $\{ 1,2,3,4,5 \}$. The right vertices code the partition $\{ \{1\}, \{ 4 \}, \{2,5 \} , \{ 3 \} \} $. The surface is connected through the interior via $5$ sheets, drawn as edges. }
\label{Edges}
\end{figure}
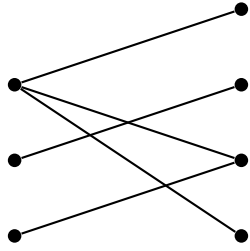

Let us first provide a glimmer of geometric 
intuition.\footnote{We recommend reading the detailed description in Example 2.1 in \cite{AN1} as well as consulting Figure 5 of \cite{Alexeevski:2002rp} for more background.} 
Consider a boundary operator insertion that maps onto a particular $n$-fold cover with boundary that allows for seamed surfaces i.e. boundaries that may be sewn together at a seam. To the left of the boundary operator, we have an $n$-fold cover of the boundary and the $n$ sheets can be divided into groups that are sewn together on the boundary. This specifies a set partition of the set of sheets $[n]=\{1,2,\dots,n \}$. On the right of the boundary insertion, we have a different set partition. The left and the right partition are connected through the bulk by the $n$ sheets. We can draw these as edges between the vertices on the left and the vertices on the right. (Each vertex represents a seam and there are as many edges departing from a vertex as there are sewn sheets at that seam.) See Figure \ref{Edges}.

All the ways in which these connections can be made, up to a relabeling of sheets, represent different boundary vertex insertions. 
The set partition on the left gives rise to a representation of $S_n$ on the left. The right partition as well. The count of the number of boundary vertex operators is the number of (group compatible) homomorphisms from this representation of $S_n$ to itself.\footnote{ This is Corollary 2.1 in \cite{AN1}. }

In our analysis, we want to be more quantitative about the number of boundary vertex operators and their representation content, given a value $n$. For $n=3$, the number of boundary vertex operators is known to be $10$ \cite{AN1} and the algebra of the boundary vertex operators was computed. In this section, we will obtain considerably more information about the boundary vertex algebra for other values of $n$, the dimension of the algebra for higher values of $n$, and its rapid asymptotic growth.

\subsubsection{The Decomposition of the Bell Representation}
To understand the boundary operator algebra, it is sufficient to understand the symmetric group representation on the partitioned left vertices. Consider then the representation of the symmetric group on the set of set partitions of $\{1,2,\dots,n \}$. There are $B_n$ such set partitions where $B_n$ is the Bell number. We call the corresponding representation the Bell representation. Once we understand the Bell representation, the group compatible homomorphisms of this representation to itself will constitute the boundary algebra. Because of Schur's lemma, the algebra of homomorphisms will equal a direct sum of square matrix algebras of sizes equal to the multiplicity of each irreducible representation (squared, and summed over all irreducibles). 

In other words, we denote by $\Pi_n$ the set of set partitions of $\{1,\ldots,n\}$ and by $V_n$ the permutation representation of $S_n$ on the vector space with basis $\Pi_n$. We clearly have $
\dim V_n=|\Pi_n|=B_n \, . $ We want to determine the decomposition
\begin{equation}
V_n\cong \bigoplus_{\lambda\vdash n}m_\lambda R^\lambda \, ,
\end{equation}
where $R^\lambda$ is the irreducible representation $\lambda$. The algebra of group compatible homomorphisms then equals
\begin{equation}
{\End}_{S_n}(V_n) \cong \bigoplus_{\lambda\vdash n} \text{Mat}_{m_\lambda}(\mathbb C) \, ,
\end{equation}
by Schur's lemma. Hence
\begin{equation}
\dim {\End}_{S_n}(V_n) = \sum_{\lambda\vdash n}m_\lambda^2 \, .
\end{equation}

Let us do the calculations in a pedestrian fashion for a few low values of $n$ to get our feet wet. For $n=1$, we find the trivial representation of $S_1$ and a one-dimensional matrix algebra. For $n=2$, we have the partitions $\{ \{ \{1,2 \} \}, \{ \{1 \} ,\{ 2 \} \} \}$ i.e. sewn or un-sewn, and $S_2$ is represented in terms of $2$ trivial representations. We therefore have a matrix algebra of dimension $2^2=4$ as the boundary operator algebra.

For $n=3$, we slightly raise the level of abstraction. The set partitions of $[3]=\{1,2,3\}$ are
\begin{equation}
123,\qquad 12|3,\qquad 13|2,\qquad 23|1,\qquad 1|2|3 \, .
\end{equation}
We have $\dim V_3=5$. We can determine the multiplicities of the irreducible representations by calculating the character of the representation. We need the character evaluated on an element of each conjugacy class. The conjugacy classes of $S_3$ can be represented by
\begin{equation}
1,\qquad (12),\qquad (123) \, .
\end{equation}
Since we have a permutation representation, the character counts the number of basis vectors that are fixed by the permutation. Clearly, for the identity class we have
\begin{equation}
\chi(1)=5 \, .
\end{equation}
For a transposition, we find:
\begin{equation}
\chi((12))=3 \, ,
\end{equation}
namely three of the five partitions are invariant under a given transposition. 
For a $3$-cycle, we have
\begin{equation}
\chi((123))=2 \, .
\end{equation}
Therefore
\begin{equation}
\chi_{V_3}=(5,3,2) \, .
\end{equation}
Using the characters of the irreducible representations of $S_3$ and the orthogonality relations, 
\begin{equation}
m_\lambda = \frac1{n!} \sum_{g \in S_n} \chi_{V_n}(g)\chi^\lambda(g) \, ,
\end{equation}
we obtain the multiplicities:
\begin{equation}
m_{[3]}=3,\qquad m_{[2,1]}=1,\qquad m_{[1,1,1]}=0 \, .
\end{equation}
Hence
\begin{equation}
V_3\cong 3[3]\oplus[2,1] \, .
\end{equation}
The number of group compatible homomorphisms is therefore $1^2+3^2=10$  and their matrix algebra is that of three by three and one by one matrices. We have computed characters through the explicit representation and counting of fixed points and then used the character decomposition into irreducible characters to determine the decomposition of the representation into irreducibles. One can compare this derivation to the diagrammatic description in \cite{AN1}.

With a bit more effort or programming, one finds for $n=4$:
\begin{equation}
V_4
\cong
5[4]\oplus2[3,1]\oplus2[2,2] \, ,
\end{equation}
with dimension $5^2+2^2+2^2=33$ for the boundary algebra. For $n=5$, one obtains:
\begin{equation}
V_5 \cong 7[5] \oplus 5[4,1]\oplus 4[3,2]\oplus [2,2,1] \, .
\end{equation}
The sum of the squares of the multiplicities equals 91. For each finite and low enough $n$ this can clearly be worked out algorithmically. The calculation becomes computationally costly at large $n$. 

\subsubsection{The Bipartite Multigraphs and Matrices}

We wish to relate the problem of counting boundary operators to another problem, which may be more standard, in which one enumerates matrices of a certain type, up to an equivalence relation. Consider two set partitions
\begin{equation}
\pi=\{A_1,\ldots,A_r\}, \qquad \rho=\{B_1,\ldots,B_s\} \, .
\end{equation}
We associate to each such duo of partitions the matrix
\begin{equation}
M(\pi,\rho)=(m_{ij}) \, ,
\end{equation}
where
\begin{equation}
m_{ij}=|A_i\cap B_j| \, .
\end{equation}
We can think of $m_{ij}$ as counting the edges that connect vertex $i$ on the left of our boundary operator to vertex $j$ on the right. This is a nonnegative integer matrix satisfying
\begin{equation}
\sum_{i,j}m_{ij}=n \, ,
\end{equation}
since the total number of edges equals the number of sheets $n$. 
There are no zero rows or columns because then we would not draw a vertex. Changing the labels of the blocks of $\pi$ and $\rho$ corresponds to permuting rows and columns i.e. to relabeling vertices. Relabeling the underlying set $\{1,\ldots,n\}$ leaves the matrix unchanged - the description automatically takes into account the gauge invariance under the relabeling of sheets. Consequently, the boundary operators are equivalent to the set of equivalence classes of matrices
\begin{equation}
M=(m_{ij})
\end{equation}
with
\begin{equation}
m_{ij}\ge0,\qquad \sum_{i,j}m_{ij}=n \, ,
\end{equation}
and no zero rows or columns, with an equivalence under independent row and column permutations. One conclusion is that we must have an equality between the sum of the squares of the Bell representation multiplicities
\begin{equation}
\sum_{\lambda\vdash n}m_\lambda^2
\end{equation}
and the count of these equivalence classes of matrices. 

These matrices have been counted for small $n$ and correspond to series A007716 in the database OEIS \cite{OEIS}. The series starts out with $(1,4,10,33,91,\dots)$ and we find the agreement we must. Thus, we have found a more standard reformulation of the problem of counting the total number of boundary operators. Of course, the list of multiplicities determined previously contains more  information.

\subsubsection{Asymptotics}
We found it interesting to ask what the large $n$ asymptotics of the dimension $a(n)$ of the space of boundary operators is. The dimension is publicly available up to $n=50$ from a link in the OEIS database \cite{OEIS}. The asymptotics can be derived with more effort, and it equals:
\begin{equation}
a(n) \approx \frac{B_n^2}{  n!}  \exp(\frac{\rho_n^2}{2}) \, ,
\end{equation}
where the quantity $\rho_n$ equals $\rho_n=n B_{n-1}/B_n$.\footnote{The function $\rho_n$  grows like $\log n$ at large $n$.} The formula for the logarithm of the dimension has corrections that slowly fade as a power of $\log n$ divided by $n$. One can determine that the ensemble of boundary operators is dominated by the characters of (multiple) transpositions. We provide a preliminary analysis of this interesting counting problem in Appendix \ref{AsymptoticsDegeneracy}.

\subsection{Summary}
We have provided links between the general formalism of section \ref{DefectTQFT} and various applications in the literature to stress its versatility. The combinatorics of the defects of \cite{HST} as well as of the boundary conditions of seamed surfaces \cite{AN1} provide interesting geometric applications of the broad framework of defect field theories.

\section{Conclusions}
\label{Conclusions}
We studied two-dimensional topological defect field theories with symmetric group gauge symmetries. We argued that the grand canonical description of the induction and restriction defects naturally leads to Khovanov's categorification of the Heisenberg algebra \cite{Khovanov}. The endomorphisms of the unit object in the monoidal category form the bulk algebra of partial permutations providing yet another avenue to the construction of the grand canonical bulk algebra. 

We argued that the algebra of defects needs to be augmented with descendants which naturally reproduce the full operator set that describes the equivariant Gromov-Witten theory. In turn these operators complete into a full W-algebra governing the integrability of the model. The oscillator description of correlators is the bridge between the defect field theory and the Gromov-Witten dual, which deserves further study. We also provided a bimodule and induction/restriction description of the defects in \cite{HST} and provided further data on the boundary algebra of seamed surfaces \cite{AN1,Alexeevski:2002rp}.

Many open directions remain. One that is of particular interest is to generalize the gauge/gravity duality for defects that we described here to the context of $AdS_3/CFT_2$ string theory. The equivariant interpolation between the Hurwitz theory and its filtered counterpart \cite{Troost:2026abx} should allow for this generalization. In NSNS string theory, the domain walls that we described correspond to fundamental winding strings. One aspect that may be interesting to study is whether an integrable structure arises from studying the fundamental string defect operators in $AdS_3$ string theories.

\section*{Acknowledgments} 
It is a pleasure to thank my colleagues for providing a stimulating research environment. In particular, I thank Lior Benizri and Gabriel Wong for discussions. Claude Opus 5.5 under the AI for Science program of Anthropic and ChatGPT by OpenAI assisted me throughout this work. Claude was particularly useful for identifying relevant literature, drawing figures and diagrams, the proof in  Appendix \ref{ProofCharacter} as well as for detailed calculations. I remain fully responsible for the entire content of the paper. 

\appendix

\section{The Dimension of the Boundary Algebra}
\label{AsymptoticsDegeneracy}
In this Appendix, we provide an estimate for the asymptotic growth of the dimension of the boundary algebra associated to the Bell representation of $S_n$. 
\subsection{The Counting Problem}
Let us review the counting problem posed in subsection \ref{SeamedSurfaces}. Let $\chi(\sigma)=\#\{\text{Pa}\in\Pi_n:\sigma \text{Pa}=\text{Pa}\}$ be the number of set partitions fixed by $\sigma$, i.e. the character of the Bell representation $V_n=\mathbb{C}[\Pi_n]$. We have that $\chi(1)=B_n$, the $n$-th Bell number. If we decompose the Bell character $\chi=\sum_\lambda m_\lambda\chi^\lambda$ into irreducibles, we obtain the degeneracy $a(n)$ of the boundary algebra as:
\begin{equation}\label{Burnside} a(n)  = \sum_\lambda m_\lambda^2
= \langle\chi,\chi\rangle
= \frac{1}{n!}\sum_{\sigma\in S_n}\chi(\sigma)^2
= \sum_{\lambda\vdash n}\frac{\chi(\lambda)^2}{z_\lambda} \, ,
\end{equation}
 where $z_\lambda=\prod_{\ell\ge1}\ell^{m_\ell}m_\ell!$ is the order of the centralizer of the conjugacy class of cycle type $\lambda$ with $m_\ell$ cycles of length $\ell$. We want to determine the dominant contributions to the sum at large $n$. 

\subsection{Reformulating the Character of the Bell Representation}
\label{ProofCharacter}
To determine the dominant contributions, we first reformulate the expression for the Bell character entirely. 
The character is a class function and we therefore analyze the contribution class by class. Let $\sigma\in S_n$ have cycles $C_1,\dots,C_c$ of lengths $\ell_1,\dots,\ell_c$. Then we will prove that
\begin{equation}\label{SimplerCharacterFormula}
\chi(\sigma) = \sum_{\mathcal{G}}\ \prod_{G\in\mathcal{G}}f(G),
\qquad
f(G)=\sum_{d\,\mid\,\gcd\{\ell_i\,:\,i\in G\}} d^{\,|G|-1} \, ,
\end{equation}
where the outer sum runs over all set partitions $\mathcal{G}$ of the set of cycles $\{C_1,\dots,C_c\}$.

We prove this formula by counting invariant set partitions.
Let $\text{Pa}$ be a $\sigma$-invariant set partition. Let $B\in \text{Pa}$ be a block in the partition that satisfies $\sigma^dB=B$ with $d$ minimal. The disjoint union $U=B\cup\sigma B\cup\dots\cup\sigma^{d-1}B$ is $\sigma$-invariant and therefore is a union of $\sigma$ cycles. Moreover, every cycle that is contained in $U$ must intersect the set $B$ (since otherwise, it would not be a subset of $U$, generated from $B$ by $\sigma$). 

Consider the action of $\sigma^d$ on a cycle $C$ of $\sigma$. We have that $d$ must divide the length $\ell_C$ of the cycle because all the sets $\sigma^i B$ for $0 \le i \le d-1$ are pairwise disjoint. We have that $B \cap C$ is a $\sigma^d$ orbit of size $\ell_C/d$. As an example, one can consider a block $\{1,4 \}$ and a permutation $(123456)$. The cycle is of length $6$ and $d=3$.

Thus each $\sigma$-orbit of blocks determines a group $G$ of cycles, namely the cycles in $U$. Clearly, the groups we obtain in this manner for all blocks will in the end partition the set of cycles $\{C_1,\dots,C_c \}$ of $\sigma$. Suppose we have such a group $G$ of size $|G|=m$. Then the admissible orbit sizes for the blocks are exactly the divisors $d$ of the greatest common divisor $\gcd (\ell_i)$ of the lengths $\ell_i \in G$, since $d$ needs to divide all of the lengths of the cycles in $G$. For each choice of such a divisor $d$, the orbit is determined by choosing one of the $d$ orbits of $\sigma^d$ on each of the $m$ cycles of $G$, i.e. $d^{\,m}$ choices, of which $d$ give the same orbit of blocks, by translation by $\sigma$ on the blocks. This yields $d^{\,m-1}$ block orbits. That implies the claimed result (\ref{SimplerCharacterFormula}).

Let us give two further simple examples to illustrate the proof. Consider the permutation $\sigma=(12)(34)$ for $n=4$. We can put both cycles in one group $G$ or in two. In the first case, we have $m=2$. We can have $d=1$ for $\{ 1,2,3,4 \}$ or $d=2$ for $13|24$ and $14|23$. The contribution to the quantity $f$ in equation (\ref{SimplerCharacterFormula}) is $3$ for this choice of cycle set partition. For the other choice, we have $m=1$ for both groups. We then sum over $d=1,2$ and find $f=1+1$. We do this for both groups, and find a contribution $2 \times 2=4$. We have a total of $3+4=7$. One can enumerate the 7 invariant set partitions of $\{ 1,2,3,4 \}$ that confirm this result. 

The second example is $\sigma=(1234)(56)$. If we have one group $G$ then $m=2$ again and we have contributions $f=1^1+2^1=3$, while for two groups we have $f_1=3$ (for the three divisors of $4$) multiplied by $f_2=2$ for a total of $6$. The grand total is $9$. To check, one enumerates set partitions invariant under $\sigma$ (including e.g. $135|246$) and again confirms the result.

\subsection{The Asymptotics}
In the following, we determine an asymptotic approximation to the dimension of the boundary algebra. It has its use since the dimension $a(n)$ is costly to compute exactly at values of $n$ above $100$.  
\subsubsection{Identifying Terms}

Taking only the $d=1$ terms of every $f(G)$ (which contributes $1$ to $f(G)$) gives all partitions whose blocks are individually fixed by $\sigma$. These partitions are in bijection with set partitions of the set of cycles of $\sigma$. Hence
\begin{equation}\label{chiapprox}
\chi(\lambda)=B_{c(\lambda)}+ (\text{terms needing a group with }\gcd\ge2 ) \, ,
\end{equation}
where $c(\lambda)$ denotes the number of cycles in the conjugacy class $[\sigma]=\lambda$. In the following, we concentrate on the first term and we argue later that this is the dominant term. A first special case is the character of the identity class which is 
\begin{align}
\chi(1^n)&=B_n \, .
\end{align}
We can divide (\ref{Burnside}) by this identity term and define the remaining factor $R(n)$:
\begin{equation}
a(n)=\frac{B_n^{\,2}}{n!}\,R(n),
\qquad
R(n)=\sum_{\lambda\vdash n} (\frac{\chi(\lambda)}{B_n} )^{2}\frac{n!}{z_\lambda} \, .
\end{equation}
Every term in the factor $R(n)$ is positive and the identity class contributes a $1$ to the factor. The factor $B_n^2/n!$ is the naive estimate in which we square the number of set partitions and divide by the order of the gauge group. The remainder $R(n)$ codes the enhancement due to non-trivial conjugacy classes. We must wonder which classes contribute dominantly at large $n$.

\subsubsection{Determining the Dominant Classes}
To perform our analysis of the asymptotics, we introduce notations for the ratios of Bell numbers:
\begin{equation}
\beta_n =\frac{B_{n-1}}{B_n},\qquad \rho_n =n\beta_n=\frac{nB_{n-1}}{B_n} \, ,
\end{equation}
which are useful in estimating the order of various contributions to the Bell character. A useful fact is that the renormalized ratio $\rho_n$ grows logarithmically with $n$. 
This is because we have the definition and estimates \cite{MoserWyman}:
\begin{align}
e^{W(n)} &= \frac{n}{W(n)}
\nonumber \\
\frac{B_n}{B_{n-1}}  &= e^{W(n)} + O(\frac{1}{W(n)})
\nonumber \\
W(n) &= \log n + O(\log \log n) \, ,
\end{align}
where $W$ is the Lambert function. Given these estimates, let us compare the contributions of classes $\lambda$ with a single 
 $\ell$-cycle ($\ell\ge2$) and $n-\ell$ fixed points. The size of this conjugacy class in the group $S_n$ is $\binom{n}{\ell}(\ell-1)!\approx n^\ell/\ell$, and if we ignore higher greatest common divisors as in (\ref{chiapprox}), we estimate $\chi/B_n\approx B_{n-\ell+1}/B_n\approx\beta_n^{\,\ell-1}$. The contribution of this term to the factor $R(n)$ is therefore approximated by
\begin{equation}
\label{Estimate}
\frac{n^{\ell}}{\ell}\,\beta_n^{\,2(\ell-1)}
=\frac{1}{\ell}\,\frac{(n\beta_n)^{2\ell-2}}{n^{\ell-2}}
=\frac{\rho_n^{\,2\ell-2}}{\ell\,n^{\ell-2}} \, .
\end{equation}
Because $\rho_n$ grows only like $\log n$, every $\ell\ge3$ contribution tends to $0$ and only transpositions survive in the limit. The size of the conjugacy class grows like $n^\ell$ while the squared character is suppressed by $2(\ell-1)$ powers of $n/\rho_n$ because there are fewer fixed points. Only for $\ell=2$ do these two effects compensate each other. Note that we have learned more. From the estimate (\ref{Estimate}), we conclude that the contribution at $\ell=2$ is logarithm-squared enhanced compared to the identity contribution.

\subsubsection{The Leading Exponent}
It is then natural to consider also the terms arising from $k$ transpositions. 
\begin{align}
\chi(2^k1^{n-2k})&=B_{n-k}+ \dots
\, .
\end{align}
In fact, these terms exponentiate. The class size is $n!/z_\lambda=n!/((n-2k)!2^kk!)$. Moreover
\begin{equation}
\frac{B_{n-k}}{B_n}=\prod_{j=0}^{k-1}\frac{B_{n-j-1}}{B_{n-j}}
=\prod_{j=0}^{k-1}\frac{\rho_{n-j}}{n-j}
\approx  (\frac{\rho_n}{n} )^{k} \, .
\end{equation}
We can then approximate further:
\begin{equation}
R(n) 
\approx \sum_{k\ge0}\frac{1}{k!} (\frac{\rho_n^2}{2} )^{k}
= e^{\frac{\rho_n^2}{2}} \, ,
\end{equation}
that is,
\begin{equation}
\label{Approximation}
a(n) \approx\frac{B_n^{\,2}}{n!} \, \exp(\frac{\rho_n^{\,2}}{2} ) \, .
\end{equation}
The dominant contribution comes from a number of transpositions $k \approx \rho_n^2/2$. (The dominant term in the Taylor expansion of the exponential is the $k$-th term where $k$ roughly equals the exponent.) Recall that $\rho_n$ grows logarithmically with $n$. The values of $n$ at which the approximation (\ref{Approximation}) becomes valid are very large because there will be corrections of the type $(\log n)^m/n$, with bounded $m$, which only become modest at very large $n$. 

Finally, we provide intuition for why we can neglect the terms with greatest common divisor larger than or equal to two. The basic reason is that this implies that a block is moved by $\sigma$ while permutations that matter most have as many fixed points as possible. The most elementary example is to consider one transposition $(ij)$. The elements $i$ and $j$ will either be in the same block or in different ones. In the first case, one will have $B_{n-1}$ set partitions, while in the second only $B_{n-2}$ (up to overall factors). Thus, such a $d=2$ term is subdominant by a factor of $(\log n)/n$. Note that $\rho^2/2$ transpositions could then give a correction of order $(\log n)^3/n$. Still, these are subdominant, and for these and all other corrections, the power of the logarithm is bounded. The argument remains qualitatively the same for other types of cycles. 
Considering higher common divisors $d$ in equation (\ref{SimplerCharacterFormula}) is too costly. 



\end{document}